\documentclass[11pt]{article}
\usepackage[utf8]{inputenc}
\usepackage{authblk}
\usepackage{amsmath}
\usepackage{graphicx}
\usepackage{setspace}
\usepackage{xcolor}
\usepackage{comment}
\usepackage[margin=1in]{geometry}

\usepackage{amsmath, amssymb, amsthm, amsfonts}
\usepackage{color,cleveref}
\usepackage{graphicx,subcaption} 

\begin{document}

\title{Design of resilient structures by randomization and bistability}
\author[1]{Debdeep Bhattacharya\footnote{corresponding author, bhattacharya@grinnell.edu, https://orcid.org/0000-0001-5171-5506 } }
\author[2]{Tyler P. Evans}
\author[2]{Andrej Cherkaev}
\affil[1]{Department of Mathematics, Grinnell College, Grinnell, IA 50112, United States of America}
\affil[2]{Department of Mathematics, University of Utah, Salt Lake City, UT 84112, United States of America}

\maketitle

\abstract{This paper examines various ways of improving the impact resilience of protective structures. Such structures' purpose is to dissipate an impact's energy while avoiding cracking and failure. We have tested the reaction of plane elastic-brittle lattices to an impulse. Four topologies are compared: periodic triangular, square, and hexagonal topologies, and aperiodic Penrose topology. Then, structures with random variations of the links' stiffness, node positions, and random holes are compared. Combinations of these random factors are also considered, as well as the resilience of bistable elastic-brittle lattices with sacrificial links. Several parameters are introduced to measure the structural resilience of the compared designs: (i) the amount of dissipated impact energy, (ii) the size of broken clusters of links, and (iii) the spread of damage. The results suggest new routes for rationally designing protective structures using nonperiodic topology, bistability, and structural randomness. In particular, we find that some quantities of interest can be maximized by tuning the randomized design appropriately--- for example, randomly removing 8\% of links maximizes energy dissipation. We also find that randomization of bistable lattices can offer superior energy dissipation while reducing the connectivity between broken clusters of links.
\vspace{.5cm}

}

%\abstract{Protective structures are designed to dissipate the energy of an impact, or to redirect it away from an important asset. In the present work, 2D elasto-brittle lattices with four different topologies (triangular, square, hexagonal and Penrose) are subjected to impacts. It is shown that energy dissipation and some simple measures of structural integrity are improved by the use of an aperiodic Penrose lattice. Improvements also result from randomization of local topology and mechanical properties. Combinations of these randomizations are briefly considered, as are the application of these techniques to bistable elasto-brittle lattices. The results suggest a new route for rational design of protective structures: optimization by induced randomness.}

\tableofcontents

\section{Introduction}
The problem of resilient design has attracted significant attention \cite{cherkaev-et-al-PRSA-2023-optimal-structures, cherkaev-ryvkin-AAM-2019, ajdari-et-al-IJSS-2012,cherkaev-cherkaev-JOE-2003, cherkaev-cherkaev-CAS-2008,cherkaev-et-al-JOPCS-2011,cherkaev-leelavanichkul-IJODM-2012,diaaz-kikuchi-IJNME-1992,lipperman-et-al-JMMS-2009,soto-IJOC-2004}.
\color{black}
Protective structures, designed to shield and safeguard vital assets, operate on a fundamental principle: they must both accumulate and dissipate the energy imparted by an impact \cite{kochmann-et-al-AMR-2017,soto-IJOC-2004}.
\begin{comment}
\color{red}
\begin{itemize}
    \item D.M. Kochmann, K. Bertoldi, Exploiting microstructural instabilities in
solids and structures: from metamaterials to structural transitions, Appl.
Mech. Rev. 69 (2017) 050801, http://dx.doi.org/10.1115/1.4037966. ? )
\item Soto, CA., Structural topology optimization for crashworthiness Interna-
tional journal of crashworthiness 9 (3), 277-283
\end{itemize}
\color{black}
\end{comment}
These structures should absorb energy without failing catastrophically. Energy accumulation necessarily results in local damage within the structure. Theoretically, the material could absorb the energy until it melts,  but structures fail long before that due to stress concentration \cite{cherkaev-et-al-JOPCS-2011}. This concentration occurs around microscopic faults or imperfections within the material, leading to an inhomogeneous stress distribution \cite{cherkaev-et-al-JOPCS-2011}. 

Stress concentration is an inherent feature of typical, unstructured solid materials. Even a minute fault can cause significant inhomogeneity in a homogeneously loaded sample. This inhomogeneity leads to a localized increase in stress around the fault, exacerbating the damage and promoting crack formation. Once a crack forms, it propagates, causing the material to fail \cite{ostoja-starzweski-et-al-EFM-1997,ostoja-starzweski-IJSS-1998,braun-fernandez-saez-2016}. Cracks in an elastically isotropic lattice tend to occur in a preferred direction that depends on the underlying lattice topology. The failed structure leaves behind pieces that are in good condition and can absorb additional energy if appropriately integrated into a resilient design \cite{cherkaev-leelavanichkul-IJODM-2012,soto-IJOC-2004,cherkaev-et-al-JOPCS-2011}. 

The design of resilient metamaterials aims to counteract this natural tendency of stress concentration and subsequent crack propagation, leaving behind a larger amount of usable material. Engineered metamaterials should endure substantial damage by redistributing the stress throughout the structure. They are designed to allow the initiation and arrest of faults in different places without compromising the overall structural integrity. This capability enables the structure to absorb more energy and contain the spread of damage without catastrophic failure. The ability to stop the growth of faults is achieved by activating a designed reserved strength of the engineered material, as in, for example, materials containing bistable links \cite{slepyan-ayzenberg-stepanenko-JMPS-2004,cherkaev-et-al-JMMPS-2005,cherkaev-at-al-MOM-2006,cherkaev-leelavanichkul-IJODM-2012,nadkarni-et-al-PRE-2014}. 

This paper explores the intricate failure mechanisms of complex structures and the design principles that can enhance the resilience of protective structures. Numerical experiments are performed with lattices with breakable or transitional links and study various structural means to create resilience. The cascades of links' damage are localized in space and time and do not require continuity, making such lattices particularly suitable for such a study. For the first time, we show that crack propagation can be arrested by randomization and the use of aperiodic lattice topologies, thereby destroying the preferred direction of propagation in an isotropic lattice. The possibility of ``optimal randomization" is also suggested and will serve as the basis for forthcoming work.

The paper is organized as follows. In Section 2, we describe the model, details of simulation, and the quantities extracted from our simulations as measures of resilience. In Sections 3 and 4, we perform several numerical experiments: altering the lattice topology; perturbing the stiffnesses of links in the lattice; perturbing the locations of lattice nodes; and randomly removing links from the lattice in order to uniformly reallocate their mass. Then, in Section 5, we consider lattices with sacrificial and waiting links, and how these lattices interact with the randomization of material properties described above. Results are summarized and discussed in Section 6.

%In the next section, we formulate the problem of dynamics and cascade of breaks in the basic triangular lattice with elasto-brittle links. Then, we explore and discuss the influence of the lattice structure: its periodicity, connectedness of links (topology), random perturbations to lattice properties (such as the stress-strain relation and initial elongations), and the use of sacrificial (waiting) links. Our findings suggest a role for the strategic use of symmetry-breaking in lattice structures to improve resilience across several lattice types. %This work is presented on a ``semi-quantitative" basis; we do not attempt a mathematical treatment of the optimal design problem, as a fully quantitative criterion for impact resilience is still a matter of ongoing debate and discussion; see, for example, \cite{cherkaev-leelavanichkul-IJOES-2012}.

\section{Model}
\label{section:model}
\subsection{Elasto-brittle links} 
The initial design used for comparison is a periodic lattice with elasto-brittle links. For each link, the stress $s$ in a link linearly increases with elongation or contraction $e$ until it reaches a threshold, beyond which the link breaks. This is given by
\begin{equation}
s= k\, e \, (1- d)
\end{equation}
where $k$ is the stiffness of strength of the link. For each link, $d(t)$ is the damage indicator, which jumps from zero to one the first time $t'$ when the elongation exceeds the threshold $e_{\max} $: 
\begin{eqnarray}
& & |e(t)| \le e_{\max} ~~\forall t \in [0, t'),    \quad |e(t')|= e_{\max } \label{discretebreakage1}
\\
& &  d(t) = 
\left\{
\begin{array}{cc}
0,   &  t< t'   \\
1  &  t \geq  t' 
\end{array}
\right. \label{discretebreakage2}
\end{eqnarray} 
These and other phase-changing materials tend to form transition waves on impact \cite{cherkaev-et-al-JMMPS-2005,cherkaev-at-al-MOM-2006,nadkarni-et-al-PRE-2014,slepyan-ayzenberg-stepanenko-JMPS-2004,slepyan-et-al-JMPS-2005,vainchtein-et-al-PRB-2009}. These waves of phase transition tend to delocalize the energy of an impact, thereby promoting structural resilience \cite{kochmann-et-al-AMR-2017,soto-IJOC-2004}. There has also been recent interest in the possibility of creating tuneable and ``reprogrammable" metamaterials based on phase-changing links \cite{ajdari-et-al-IJSS-2012, kochmann-et-al-AMR-2017, francois-et-al-IJSS-2017,zareei-et-al-PNAS-2020,ramakrishnan-et-al-JAP-2020,melancon-AFM-2022}. ``Honeycomb" (hexagonal) arrays have been investigated for their resilience \cite{caccese-et-al-CS-2013,mousanezhad-et-al-IJMS-2014,mohammadi-et-al-JMRT-2023}. 
%We note briefly that, as an alternative, we could have accounted for damage by a monotonically-increasing damage function $d(t)$ satisfying the differential equation
%\begin{eqnarray}
% d(0)=0, \quad  \dot{d} =
%\left\{
%\begin{array}{cc}
%0,   &  t< t'   ~\rm{or}~ d=1\\
%V  &   \mbox{ otherwise}
%\end{array}
%\right. \label{continuousdamage}
%\end{eqnarray}
%where $V$ is the rate of damage, as in \cite{cherkaev-leelavanichkul-IJOES-2012}. We choose the former approach of Equations (\ref{discretebreakage1}) and (\ref{discretebreakage2}), and leave the latter, of Equation (\ref{continuousdamage}), for future work.

\paragraph{Energy loss in a damaged link.} Consider the motion of a single damageable link with the attached mass $m$. 
The equation of the motion is
$$
m \ddot{e} + k\,(1-d)  e=0 
$$
where $e$ is the elongation. 
Multiply by $\dot{e}$ and integrate along the trajectory to obtain
$$
\int_{t_0}^{t_1}\left( m \ddot{e} \dot{e} + k\,  e \dot{e} \right) dt=\int_{t_0}^{t_1}d\, e \dot{e} \, dt.
$$
Using the identities 
$$ \ddot{e} \dot{e}= \frac{1}{2} \frac{d}{dt} \dot{e} ^2, \quad  e \dot{e}= \frac{1}{2} \frac{d}{dt} e^2
$$
rewrite
$$
 \left. \frac{1}{2} \left( m\dot{e} ^2 + k \,e^2 \right) \right|_{e(t_0)}^{e(t_1)}
=\frac{k}{2}
\int_{t_0}^{t_1}  d \left(  \frac{d}{dt}{e^2} \right) dt. 
$$
The left-hand side  shows the change of the whole energy of the link, and the right-hand side
%When elongation $e(t)$ makes a loop,  ${e(t_0)}={e(t_1)}$  the value of $I$  
$$ 
E_{\text{loss}}=\left\{
\begin{array}{cl}
 0 &    t>t' \cr
- \frac{k}{2}(e^2_{\max} + e(t_0) ) &  t<t' 
\end{array}
\right.
$$
shows the energy loss due to the link breakage. In our experiments, the lattice is initially not loaded, so that $e(t_0)=0$, and the lost energy $E_{\text{loss}}$ in a broken link equals the potential energy stored in that link before breakage:
\begin{equation}
E_{\text{loss}} = - \frac{k}{2}\,e^2_{\max} \label{energylost}
\end{equation}

\subsection{Dynamics}
\label{section:dynamics}
\begin{figure}[htpb]
\centering
    \subfloat[]{
        \includegraphics[width=0.18\linewidth, clip=true, trim={40 0 60 0}]{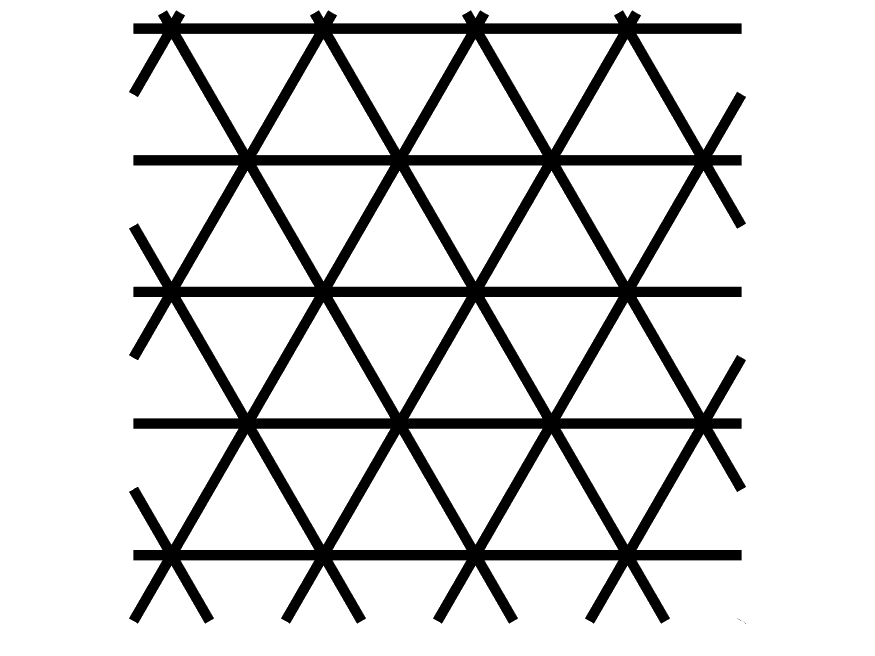}
    }
    \subfloat[]{
        \includegraphics[width=0.18\linewidth, clip=true, trim={40 0 60 0}]{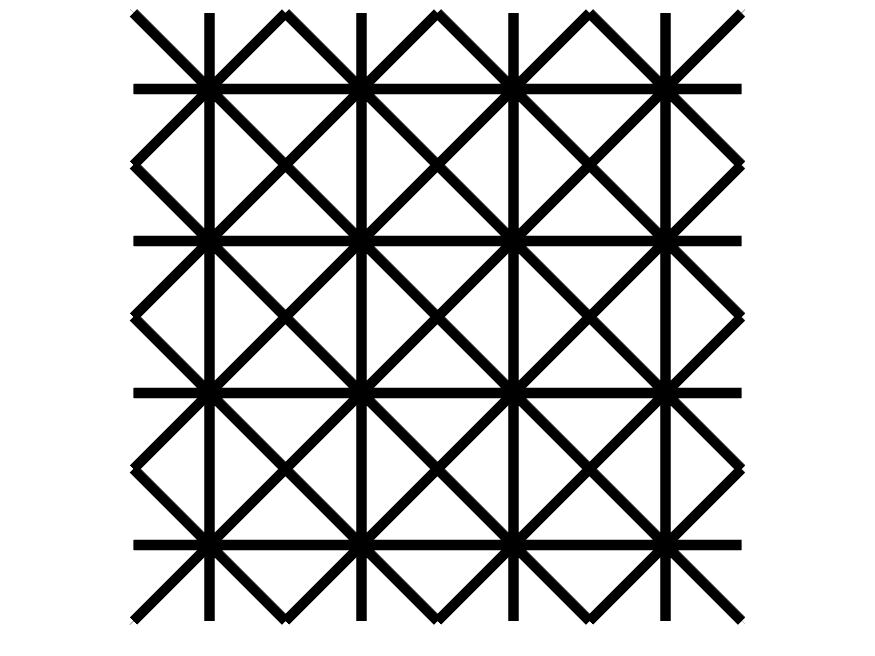}
    }
    \subfloat[]{
        \includegraphics[width=0.18\linewidth, clip=true, trim={40 0 60 0}]{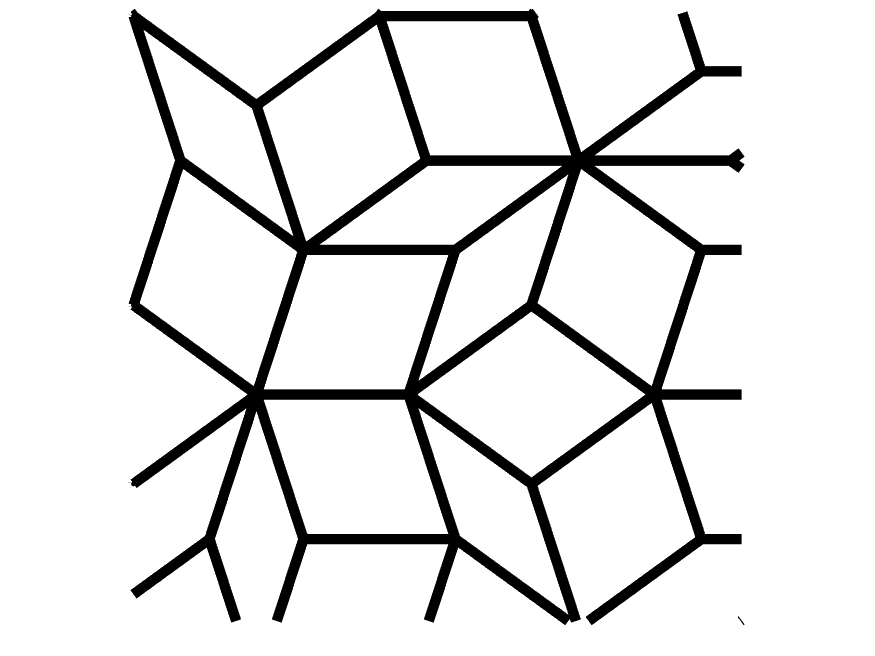}
    }
    \subfloat[]{
        \includegraphics[width=0.18\linewidth, clip=true, trim={40 0 60 0}]{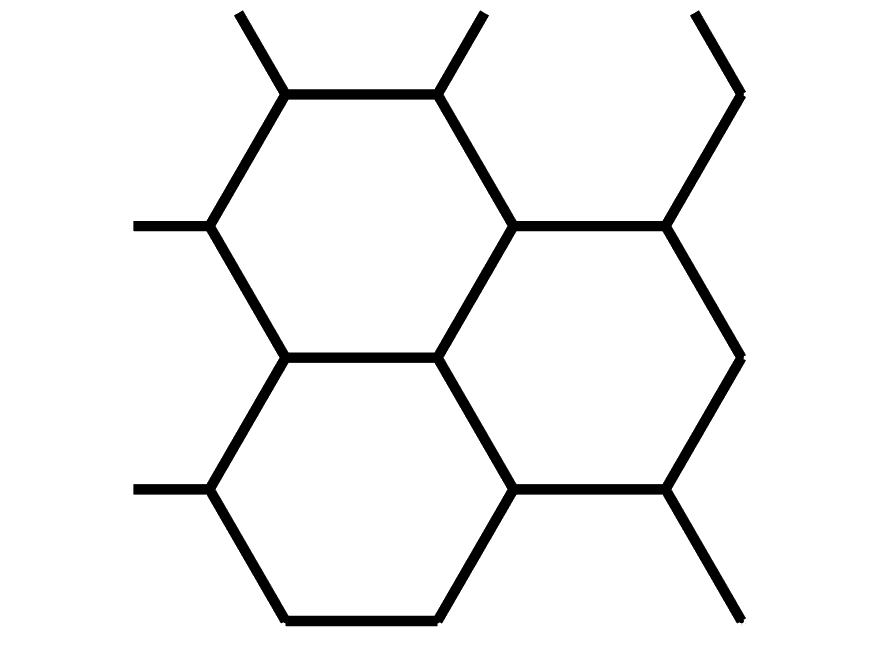}
    }
\caption{A few cells of the four lattice topologies relevant to this work: (a) triangular lattice; (b) square lattice with diagonals; (c) aperiodic Penrose lattice; and (d) hexagonal or ``honeycomb" lattice. }%
\label{fig:topologies}
\end{figure}

In this work, we consider four lattice topologies: a periodic triangular lattice, a periodic square lattice, an aperiodic Penrose lattice and a hexagonal lattice. A few cells of each lattice are shown in Figure \ref{fig:topologies}.

Each lattice is modeled as a set of massless links $l_{ij}$ joining nodes $x_{i}$ and $x_{j}$. Each node has dimensionless mass $m$ and is connected to its nearest neighbors. Instantaneously, the nodes move according to equations
\begin{equation}
m \ddot{x}_{i}= \sum_{j \in \Omega_{i}} \kappa d_{ij} e_{ij}( x_{i}- x_{j}) + F_{i}(x_{i})
\end{equation}
where $m$ is the mass of the given node (taken to be uniform across all nodes), $x_{i}$ denotes its position in the lattice, $\Omega_{i}$ is the neighborhood of the position $x_{i}$, $d_{ij}$ indicates the state of damage in a given link (0 for an intact link, or 1 for a broken link), $\kappa$ is the stiffness of the link, and $F_{i}$ is external forcing. The whole energy $W$ (Hamiltonian) of the system is
\begin{equation}
W=\frac{1}{2} m  \sum_{i} \dot{x}^2_{i} +\frac{1}{2} \kappa \sum_{i} \bigg( \sum_{j \in \Omega_{i}} (1- d_{ij}) e_{ij}( x_{i}- x_{j})^2 \bigg),
\end{equation}
where the first term is the total kinetic energy, and the second term is the total potential energy. The energy $W$ stays constant between the instances of variation of $d_{ij}$; the system is linear and conservative in these intervals. Thus, the dynamics of a breakable lattice is described as a cascade of linear systems with decreasing energy.

\paragraph{Initial impulse.} Initially, all links are in equilibrium position with zero velocity. The lattice is then subjected to an impulse at its center, simulated by applying an outward-directed body force near the center of the domain for a fixed, short duration. The body force experienced by each node is given by the Gaussian vector field
\begin{equation}
 \vec{F}(x_{i}) = C e^{- \frac{||\vec{x}_{i}||^2}{\sigma^2} }  \frac{\vec{x}_{i}}{||\vec{x}_{i}||}  , \label{impacteqn}
\end{equation}
where $\vec{x}_{i}$ is the vector from the origin to node $x_{i}$. The magnitude of the force near the center of impact approaches $C$. Throughout this work, $C=100$ and $\sigma^2 = 120$. The edges of the domain are assumed to be immobile. At the initial time, the Gaussian force described above is applied for as long as needed so that the total system energy is increased by 850 dimensionless units; then the application of force is immediately ceased, and remains zero for the remainder of the simulation. 

Since conservation of energy is important for accurately tracking breakage, we use the fourth-order Yoshida method \cite{yoshida-PLA-1990}, an explicit energy-conserving method for Hamiltonian systems. 

\paragraph{Parameters.} The model allows for variations of link properties, including length, stiffness and threshold for breakage, as well as the location, duration and orientation of impact. The impact is modeled as an initial, outward impulse; see (\ref{impacteqn}). In order to make a fair comparison of the different lattice geometries, the dimensions of each plate are fixed at 40-by-40 dimensionless units. All links are taken to be the same length of 1 unit, except for the diagonal of the square lattice, which is unavoidably $\sqrt{2}$ units. A fixed amount of mass $M$ is available for the construction, chosen to be 10000 units. Then a constant density is assumed for the links, with the stiffness $k$ of each link proportional to the cross-sectional area. When mass is evenly distributed across links of uniform length, and $k \propto M/L$, where $L$ is the total length of links used to construct the 40-by-40 plate. The constant of proportionality is taken to be $10$. All mass is taken to be concentrated at the lattice node points, treating the links as massless for the purposes of numerical simulations. In order to eliminate a parameter, 20\% elongation is assigned as the threshold for elasto-brittle links to break, corresponding to $e_{\text{max}} = 0.2$. 

In order to emphasize the dependence on lattice topology, the origin of the impulse is the center of a cell for the triangular, square, and hexagonal lattices. For the Penrose lattice, the impulse is applied at the plate's central node. This work is focused on the damage profile at a fixed time; hence there is no consideration of effects occurring after this fixed time, nor the temporal evolution of damage. These are potential topics for future work.

Only direct ``impacts", modeled here as radially-symmetric initial impulses are considered. There is no consideration of asymmetric forcing, nor exploration of parameter space beyond the set of experiments discussed below, although these are also potential topics for future work.

\begin{figure}[htpb]
\centering
    \subfloat[]{
        \includegraphics[width=0.18\linewidth, clip=true, trim={40 0 60 0}]{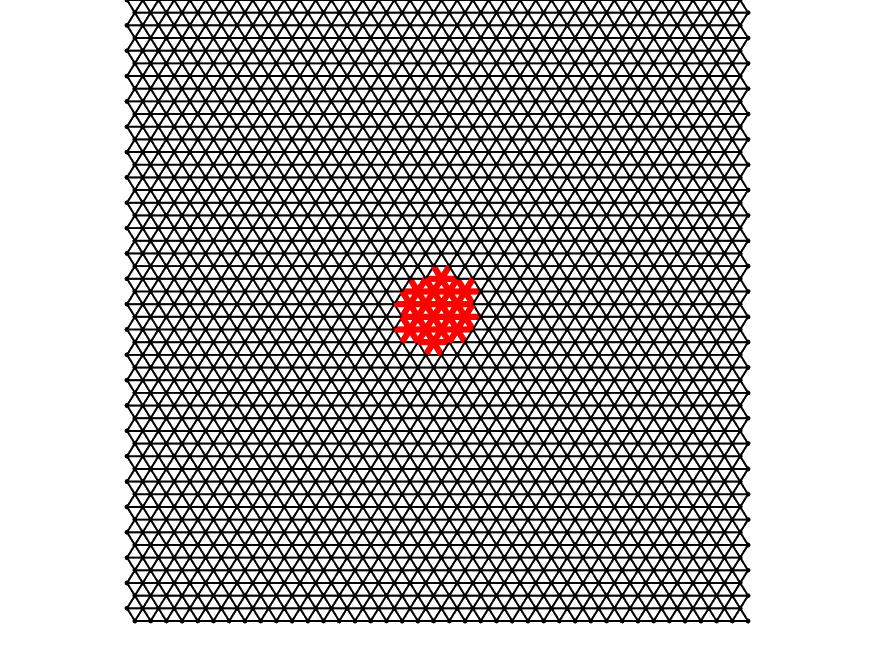}
    }
    \subfloat[]{
        \includegraphics[width=0.18\linewidth, clip=true, trim={40 0 60 0}]{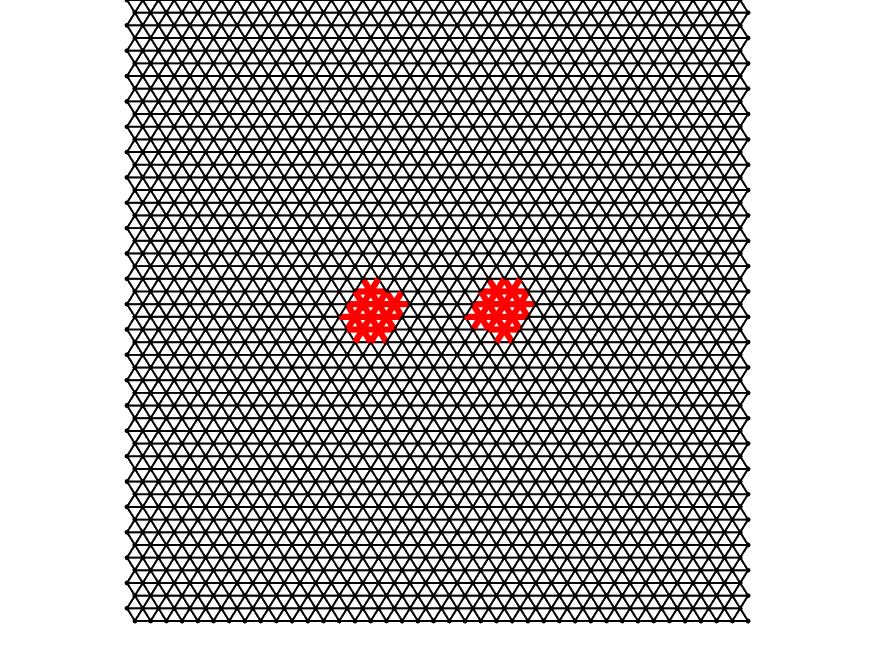}
    }
    \subfloat[]{
        \includegraphics[width=0.18\linewidth, clip=true, trim={40 0 60 0}]{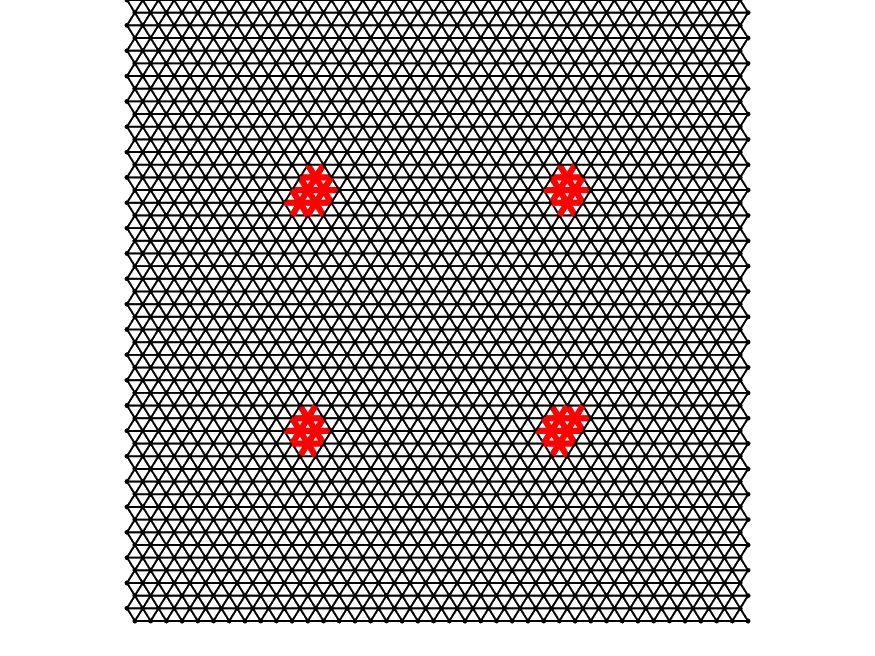}
    }
    \subfloat[]{
        \includegraphics[width=0.18\linewidth, clip=true, trim={40 0 60 0}]{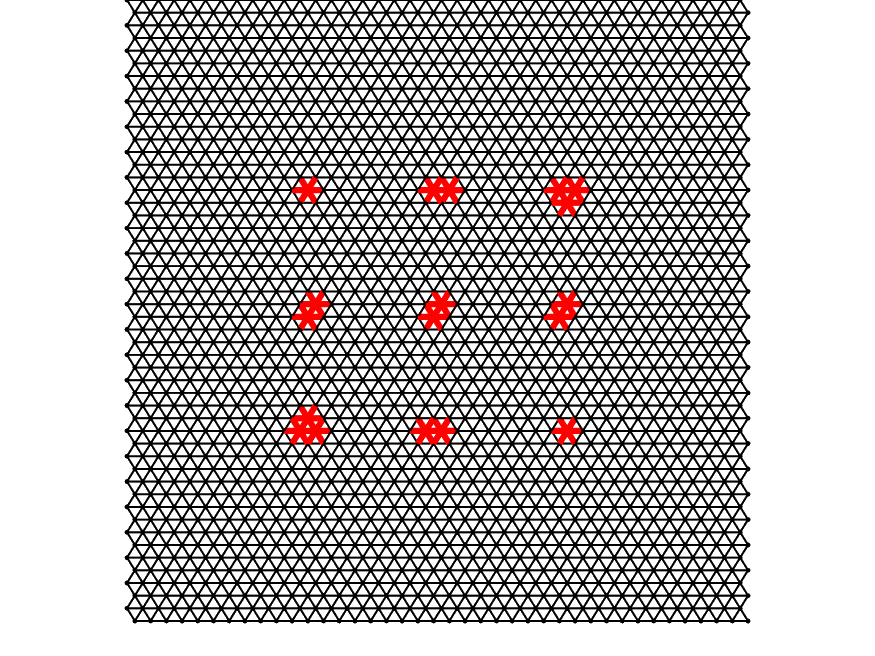}
    }
\caption{An illustration of severity measure, $S$. Each of (a)-(d) have 108 broken links (represented in red color), hence have the same amount of total dissipated energy. (a) $S=0.623$; (b) $S=0.550$; (c) $S=0.467$; (d) $S=0.367$. High $S$ is associated with concentration of damage.}%
\label{fig:severity}
\end{figure}

\paragraph{Criteria for resilience.} There is no single parameter that describes the damage in a planar lattice \cite{cherkaev-leelavanichkul-IJOES-2012}. Accordingly, we define three quantities intended to capture different aspects of the damage profile. These quantities are then collected from each numerical experiment. The following definitions adopted throughout the text:
\begin{itemize}
    \item The fraction of \textit{energy dissipated} from the initial impulse by breaking links, denoted as $D$. We define
    \begin{equation}
    D = \frac{\sum_{ij}E_{\text{loss},ij}}{E_{\text{impulse}}},
    \end{equation}
    where $E_{\text{loss},ij}$ is the dissipation of energy due to breakage of link $ij$ within the simulated time; see (\ref{energylost}). $E_{\text{impulse}}$ is the total energy added to the lattice by the initial impulse; see (\ref{impacteqn}).
    \item A simple measure of \textit{severity} of damage, denoted as $S$: the average fraction of broken links connected to a node given that at least one of its links is broken. We define
    \begin{equation}
    S = \sum_i \sum_{\substack{j \in \Omega_i \\ d_{ij} \geq 1}}\frac{d_{ij}}{|\Omega_i|}.
    \label{severity}    
    \end{equation} 
    For a fixed number of broken links, this quantity becomes large when more breaks occur around fewer nodes, suggesting more intense damage, like cracks or fractures. It becomes small when the breaks occur at many distant nodes. $|\cdot|$ is set cardinality. Figure \ref{fig:severity} visualizes $S$ for a few simple cases.
    %We denote this quantity $\mu$. %For larger $\mu$, damage tends to be more concentrated. In order to increase resilience, we prefer lower $\mu$.
    \item A measure of the \textit{spread} of dissipation: the radial distances from the center of the initial impulse so that a percentage $p$ of dissipated energy is lost within that radius. We denote these radii $r_p$. For a fixed amount of dissipated energy, these radii become large when the dissipation occurs over a larger area, signaling delocalization of impact energy. More precisely,
    \begin{equation}
    r_p = \min\bigg\{r \in \mathbb{R}^{+} : \sum_{i} \sum_{\substack{j \in \Omega_i, \\ ||\vec{x}_i|| \leq r}}  E_{\text{loss},ij} \hspace{.125cm} \geq \hspace{.125cm} p\cdot \sum_{ij}  E_{\text{loss},ij} \bigg\}.
    \end{equation}
\end{itemize}
In the above, $||\cdot ||$ is simply the Euclidean norm, and $\vec{x}_i$ is the vector that connects node $x_i$ to the origin.

By a ``resilient" structure, we mean that a large fraction of energy is dissipated, the probability that many breaks occur close together is small (low ``severity", $S$), and the energy dissipation occurs over a large area (the destruction of material is ``spread out"). While there are many possible ways of quantifying resilience for a two-dimensional lattice, we regard these three simple quantities as suitable for a first exploration of the present topic, and broadly consistent with design specifications discussed elsewhere \cite{cherkaev-cherkaev-JOE-2003,cherkaev-cherkaev-CAS-2008,cherkaev-et-al-JOPCS-2011,cherkaev-leelavanichkul-IJODM-2012}.

The above quantities, when provided throughout this paper, have been statistically averaged: Tables \ref{table:pertnodes}, \ref{table:pertk}, \ref{table:removal} and \ref{table:pertk-bistable} are the result of averaging over ten realizations of the random perturbation for each of the twenty data points plotted. This is done in order to smooth the data and ensure that the effects we report are not statistical anomalies.

\section{Lattice topology}
\label{section:topology}
\subsection{Triangular and square lattices}
\label{subsection:tri-sq}
\begin{figure}[htpb]
\centering
    \subfloat[]{
        \includegraphics[width=0.33\linewidth, clip=true, trim={40 0 60 0}]{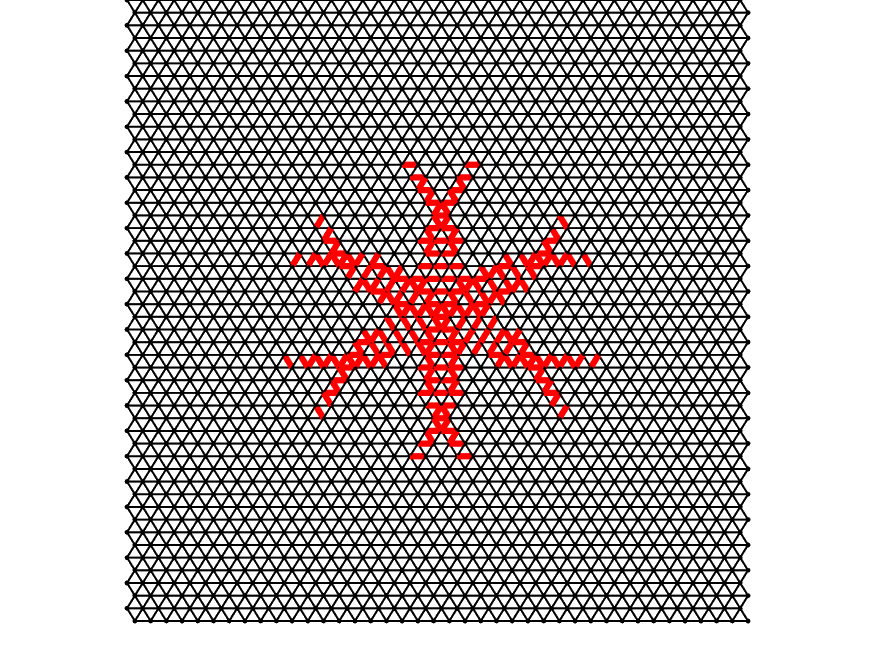}
    }
    \subfloat[]{
        \includegraphics[width=0.33\linewidth, clip=true, trim={40 0 60 0}]{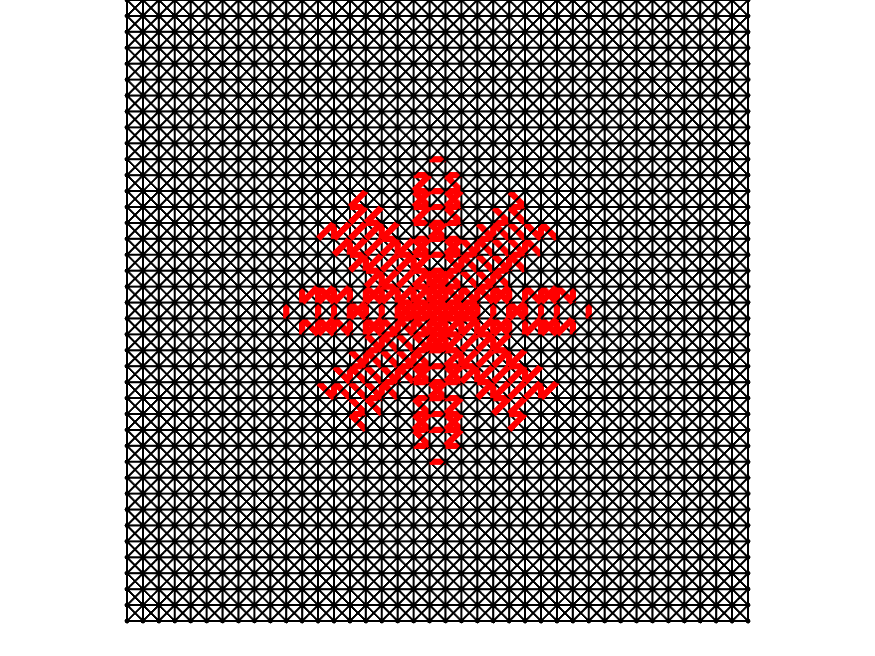}
    }
\caption{Lattice geometry and damage profile at a fixed time using the same plate area, quantity of material, energy of impact, link properties. Parameters are described in Section \ref{section:dynamics}. (a) is the triangular lattice and (b) is the square lattice with diagonals.  }%
\label{fig:shape-comp}
\end{figure}

The results of impact experiments on triangular and square lattices prior to any design interventions are shown graphically in \Cref{fig:shape-comp}; corresponding data for $D, S$ and $r_p$ are shown in \ref{table:summary-topology}. Note the shape of the crack in each of the lattice structures. It is apparent that the lattices exhibit preferred directions of crack propagation. For the triangular lattice where each node has six nearest neighbors, the resulting radially-symmetric crack has six branches that originate from the center of impact, while the square lattice has eight such branches. Notably, in contrast to the continuum case where crack propagation direction is governed by the shape of the domain boundary and 
loading profile \cite{lipton2025energy}, both the direction of crack propagation and the number of crack fronts in lattice systems are heavily influenced by the  underlying symmetry of the lattice structure.

The damaged regions in \Cref{fig:shape-comp} are highly connected: the crack has propagated along a straight path. Such cracks are associated with the catastrophic failure of the material and are undesirable for structural resilience. The ability of the crack to propagate along a straight path is evidently due to the simple translation symmetry of the lattice and the straight paths between adjacent nodes. Hence a natural question is whether lattice resilience can be improved substantially by eliminating translation symmetry.

%The dissipated energy quantifies how much energy from a projectile passes through the lattice, and the mean and standard deviations give a heuristic sense of the connectivity of the broken cluster. By ``resilience", we mean that a material can dissipate a large amount of energy (thereby protecting a valuable asset by slowing a projectile as it passes through, or rejecting the projectile completely) with as much of the original structure remaining as possible--- hence a large number of weakly-connected ``spots" of broken material are preferable over a smaller number of well-connected ``cracks" or ``holes", the presence of which suggests catastrophic failure of the material. This is broadly consistent with the notion of resilience presented in \cite{cherkaev-ryvkin-AAM-2019,cherkaev-ryvkin-AAM-2019b} and \textcolor{red}{elsewhere (MORE DETAIL HERE?)}.

\subsection{Hexagonal and Penrose lattices}
\begin{figure}[htpb]
\centering
    \subfloat[]{
        \includegraphics[width=0.33\linewidth, clip=true, trim={40 0 60 0}]{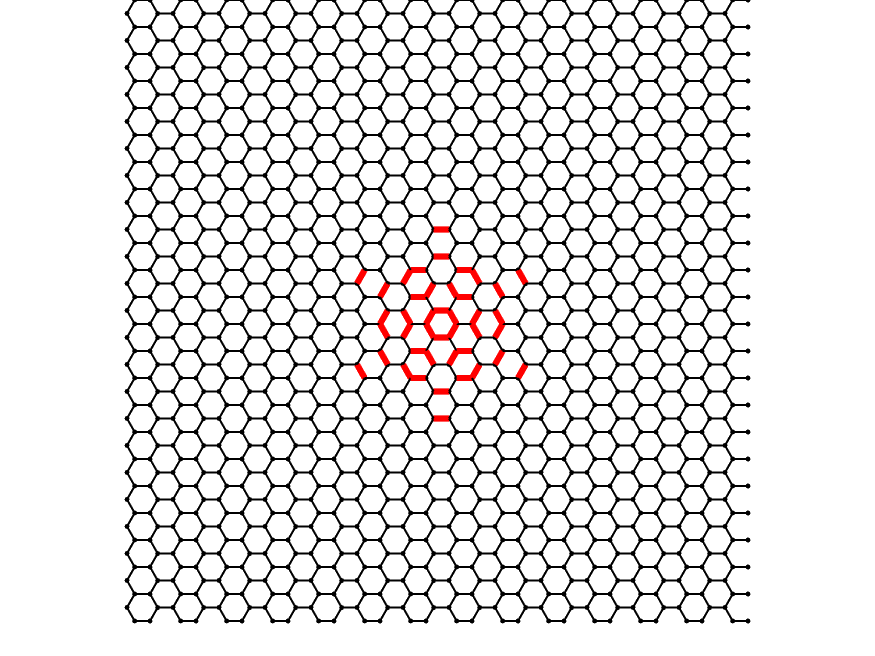}
    }
    \subfloat[]{
        \includegraphics[width=0.33\linewidth, clip=true, trim={40 0 60 0}]{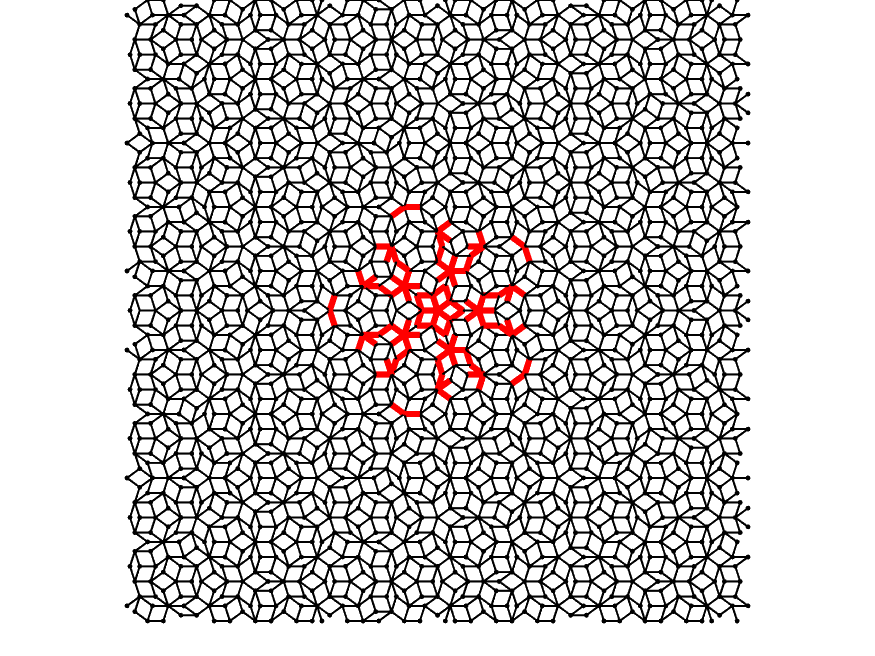}
    }
\caption{Lattice geometry and damage profile at a fixed time using the same plate area, quantity of material and impact energy. Parameters are described in Section \ref{section:dynamics}. The broken links are plotted against the reference configuration in order to emphasize the damage pattern. (a) is a hexagonal lattice and (b) is an aperiodic Penrose lattice.}%
\label{fig:shape-comp2}\end{figure}

In order to eliminate the directions of easy propagation of damage, a hexagonal lattice is considered. The results are shown graphically in Figure \ref{fig:shape-comp2}; corresponding data for $D, S$ and $r_p$ are shown in \ref{table:summary-topology}. Although still periodic, there is no single direction along which damage can easily propagate: unlike the triangular and square lattices where there exist paths along six or eight directions connecting arbitrarily distant nodes, tracing a path longer than two links in the hexagonal lattice requires a change of direction. Hence we anticipate that the hexagonal lattice is less prone to crack formation than the triangular or square lattices.

Next, an aperiodic Penrose lattice is considered. The aperiodic Penrose lattice lacks translation symmetry between adjacent cells, and nodes have three, four, or five connections, with an average of four. Based on discussion in \Cref{subsection:tri-sq}, it is expected that the Penrose lattice should also be disruptive to crack formation. To our awareness, aperiodic Penrose lattices have not yet been studied in the context of protective metamaterials.

\begin{figure}[htpb]
\centering
    \subfloat[]{
        \includegraphics[width=0.33\linewidth, clip=true, trim={40 0 60 0}]{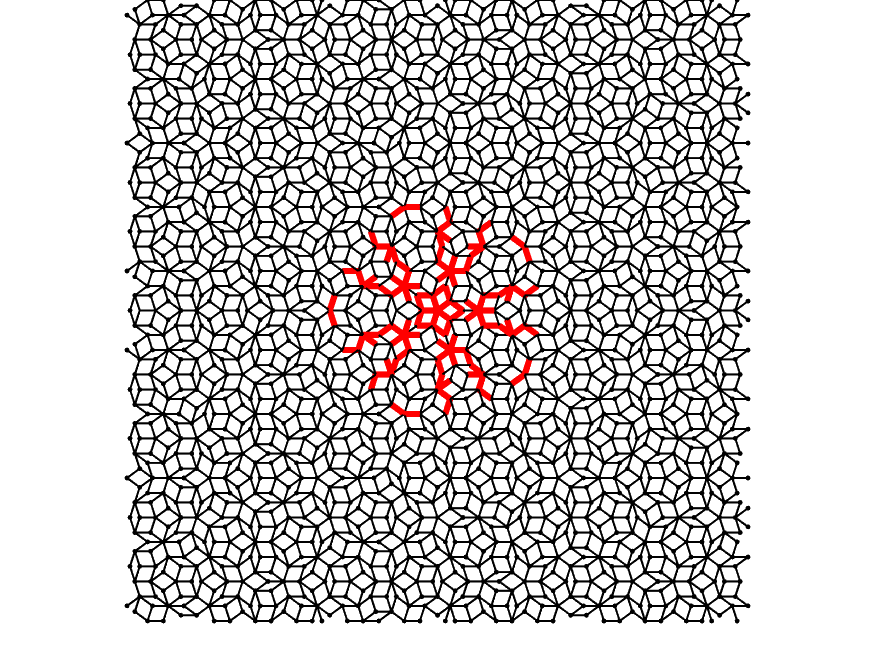}
    }
    \subfloat[]{
        \includegraphics[width=0.33\linewidth, clip=true, trim={40 0 60 0}]{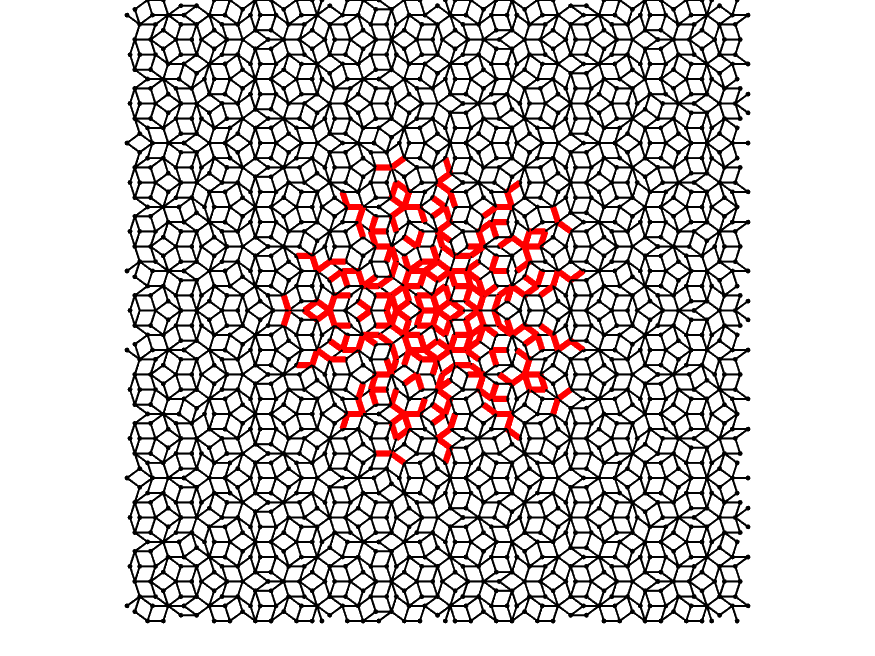}
    } \\
    \subfloat[]{
        \includegraphics[width=0.33\linewidth, clip=true, trim={40 0 60 0}]{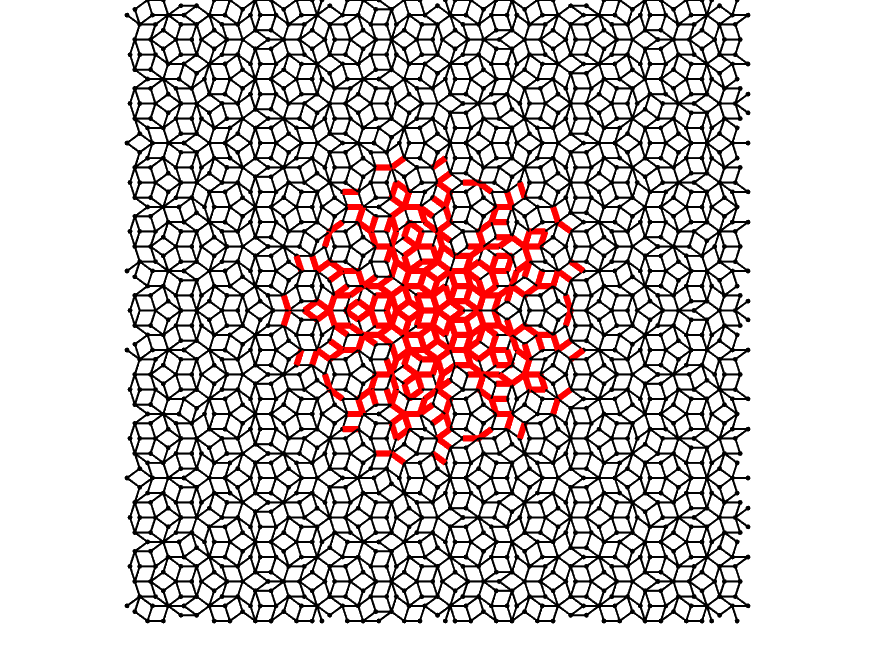}
    }
    \subfloat[]{
        \includegraphics[width=0.33\linewidth, clip=true, trim={40 0 60 0}]{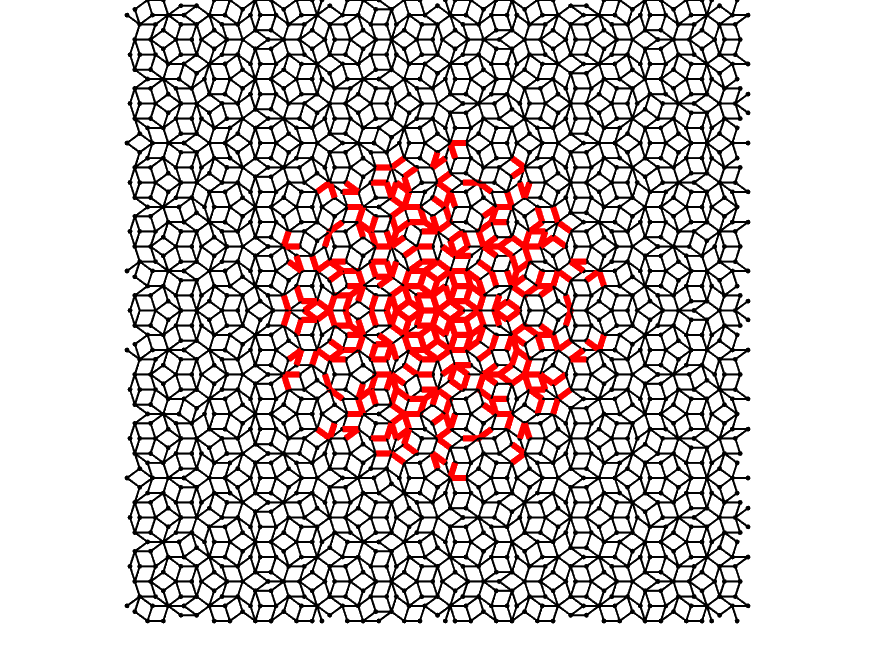}
    }
\caption{The damage pattern for the Penrose lattice for impacts adding (a) 1200, (b) 2400, (c) 3600 and (d) 4800 units of energy. Parameters are described in Section \ref{section:dynamics}. }%
\label{fig:penrose-feature}
\end{figure}

Both the hexagonal and Penrose lattices are subjected to the same initial conditions as the triangular and square lattices described above, in Equation (\ref{impacteqn}). In Figure \ref{fig:penrose-feature}(d) and Figure
\ref{fig:shape-comp2}, it can be seen that the damage in both the hexagonal and Penrose lattices is less ``severe" (resulting in lower values of $S$) than the square or triangular lattices, in the sense of (\ref{severity}). Then, in Figure \ref{fig:penrose-feature}, several results of simulated impacts on the Penrose lattice with several energies are shown. Across these figures, it can be seen that the Penrose lattice tends to disperse damage more uniformly than the triangular or square lattices: within the damaged region, there many spots of undamaged material.

\begin{figure}[htpb]
\centering
    \subfloat[]{
        \includegraphics[width=0.33\linewidth, clip=true, trim={40 0 60 0}]{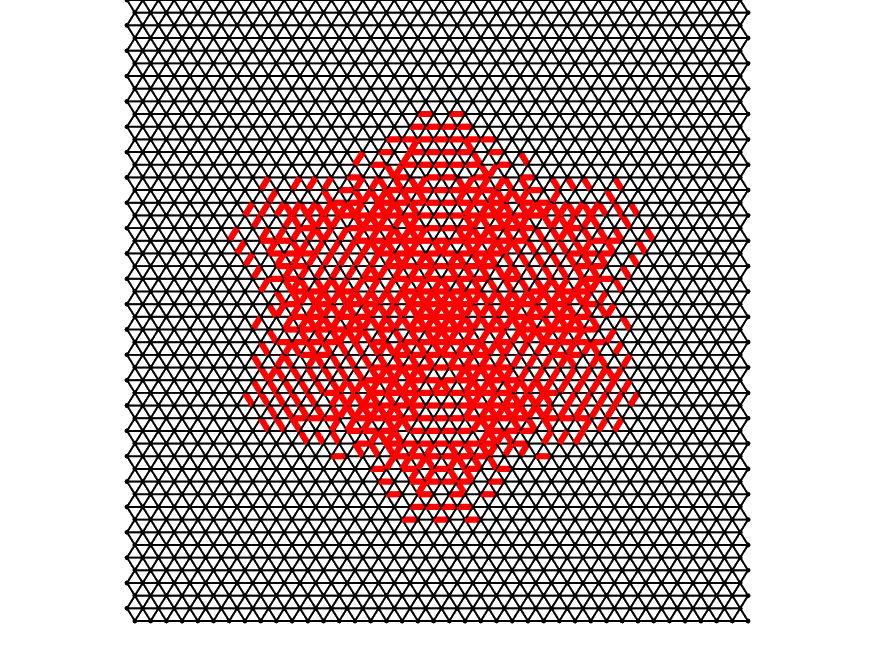}
    }
    \subfloat[]{
        \includegraphics[width=0.33\linewidth, clip=true, trim={40 0 60 0}]{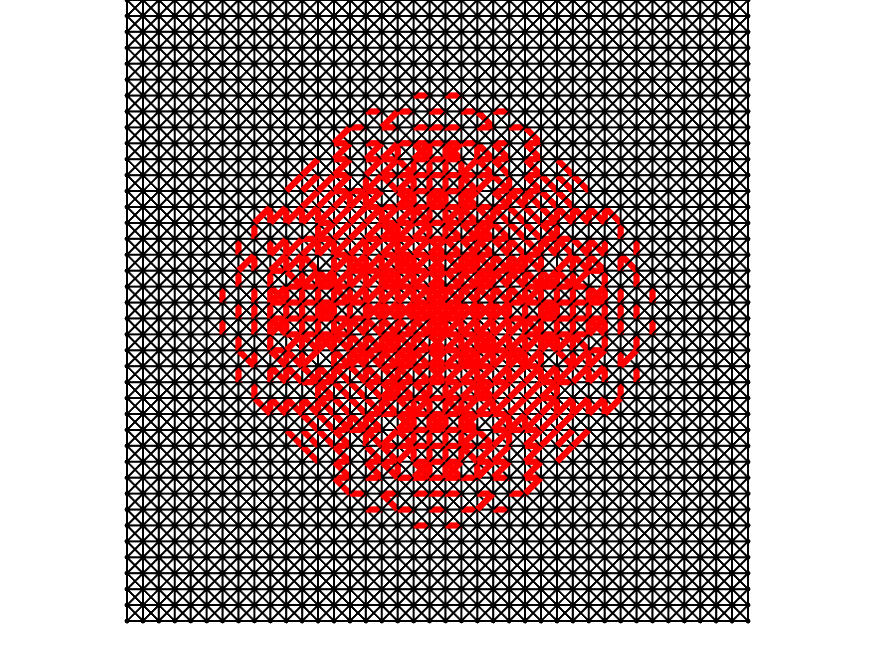}
    }
\caption{Comparison of damage patterns for triangular and square lattices at 4800 units of impact energy. Compare also with Figure \ref{fig:penrose-feature}(d). Parameters are described in Section \ref{section:dynamics}. }%
\label{fig:shape-comp3}
\end{figure}

\begin{table}[htpb]
\centering
\begin{tabular}{||c c c c c||} 
 \hline
Type & Triangular & Square & Hexagonal & Penrose \\ [0.5ex] 
 \hline\hline
 $D$ & \textit{0.097} & \textbf{0.211} & 0.056 & 0.061 \\ 
 $S\cdot \bar{|\Omega|}$ & 2.256 & 2.976 & \textbf{1.2726}  & \textit{1.676} \\
 $r_{.25}$ & 3.214 & \textbf{3.535} & 2.645 & 2.618  \\
 $r_{.50}$ & 5.131 & \textbf{5.418} & 3.605 & 4.406  \\
 $r_{.75}$ &	\textbf{7.571} & \textit{7.106} & 4.359 & 5.6261 \\
 $r_{.90}$ & \textbf{9.073} & \textit{8.514} & 6.083 & 6.6138 \\ [1ex] 
 \hline\hline
\end{tabular}
\caption{ Summary of results for initial comparison of lattice topologies as described in Section \ref{section:topology}. The best values of $D, S, r_{.25}, r_{.50}, r_{.75}$ and $r_{.90}$ are \textbf{bolded}. The second best are \textit{italicized}. In order to facilitate comparison of $S$ between lattices with different connection densities, the product $S\cdot |\bar{\Omega}|$ is used instead of $S$ alone.}
\label{table:summary-topology}
\end{table}

\paragraph{Comments on topology.} 
Different lattice topologies have been considered and the results are summarized in Table \ref{table:summary-topology}. Because comparison of damage concentration using $S$ is most natural between lattices of the same topology, Table \ref{table:summary-topology} uses $S\cdot |\bar{\Omega}|$ instead of $S$ alone, where $|\bar{\Omega}|$ is the average number of connections per node. Heuristically, this quantity expresses the number (rather than the fraction) of destroyed links around each node. The results conform with the visual impression given by Figures \ref{fig:shape-comp} and \ref{fig:shape-comp2}: that the connectivity between destroyed links for the hexagonal and Penrose lattices is relatively weak compared with that of the triangular or square lattices. Concentration of damage in the square lattices is particularly intense.

We have considered the hexagonal lattice primarily for completeness. While it exhibits good resilience in the sense of $S\cdot |\bar{\Omega}|$, the hexagonal lattice has no shear stiffness, rendering it unusable in many applications. The hexagonal lattice also dissipates the least energy of all four lattice types considered here. In contrast, the square lattice dissipates a great deal of energy, but is the most prone to crack formation and propagation.

These initial results show that local connectivity influences damage propagation resulting from a single impact, with crack formation reflecting the underlying topology. In what follows, we consider several possibilities for improving the resilience of a lattice. In particular, we utilize a \textit{minimax principle} in the style of \cite{cherkaev-cherkaev-CAS-2008}. Practical scenarios where damageable lattice materials would be implemented in order to delocalize damage may feature \textit{a priori} unknown, random, or uncertain locations of impact, and repeated impacts. Accordingly, optimal design of an impact-resistant structure should distribute damage to minimize the greatest concentration of damage in any particular location. The uncertainty of the location of the next impact suggests that a homogeneous material is required: for a fixed mass, constructions which aim to reinforce a particular subdomain of links will weaken other subdomains that become vulnerable. Therefore such a construction is not optimal; see, for example, \cite{cherkaev-ryvkin-AAM-2019b}.

\section{Randomization}
\label{section:randomization}
The results of Section \ref{section:topology} show that the triangular and square lattices dissipate energy well, but the severity $S$ of damage that they sustain is high. Motivated by the observation that hexagonal and Penrose lattices experience much lower $S$ due to a lack of easy directions of propagation, we improve the performance of periodic triangular lattices by breaking node-wise symmetry in several different ways: hence the intelligent use of disorder in metamaterials leads to an increase in resilience to impact. 

The triangular lattice with the same construction as described in Section \ref{section:model} is used as a control. Then several numerical experiments are performed in order to investigate different kinds of symmetry breaking. These are described in the following sections.

\subsection{Random perturbation of node positions} \label{section:randomization:pertnodes}
\begin{figure}[htpb]
\centering
    \subfloat[]{
        \includegraphics[width=0.33\linewidth, clip=true, trim={40 0 60 0}]{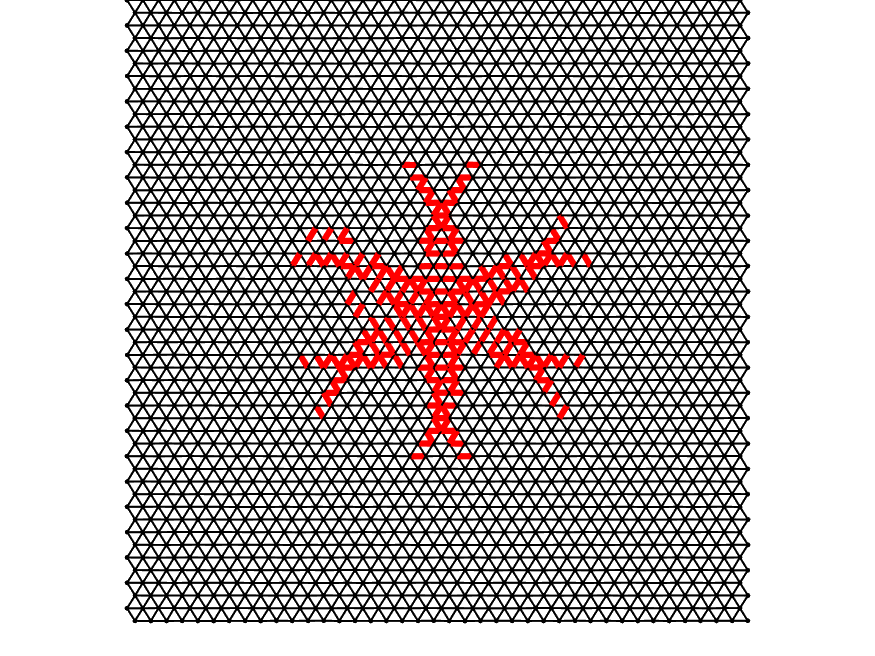}
    }
    \subfloat[]{
        \includegraphics[width=0.33\linewidth, clip=true, trim={40 0 60 0}]{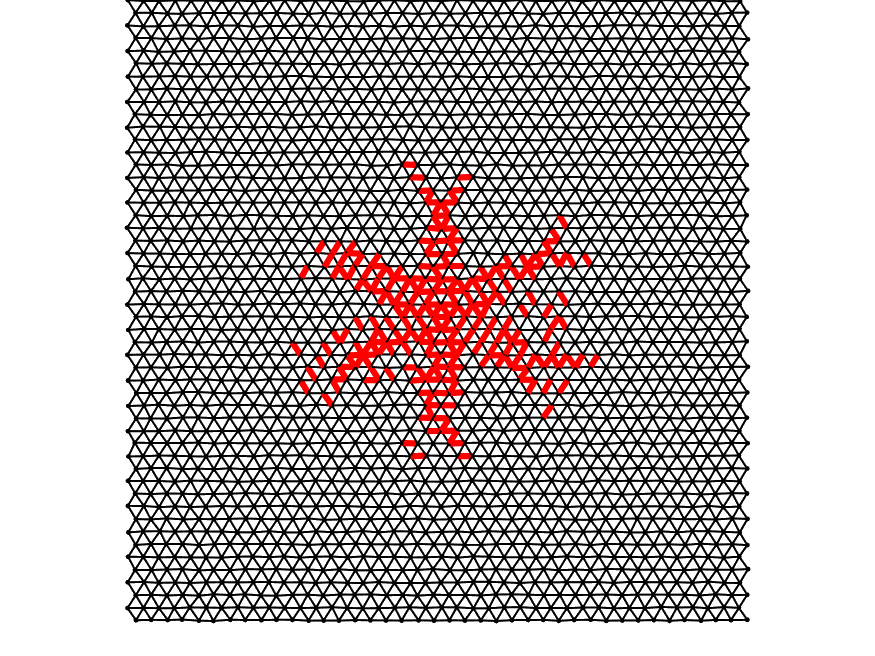}
    }
    \\
    \subfloat[]{
        \includegraphics[width=0.33\linewidth, clip=true, trim={40 0 60 0}]{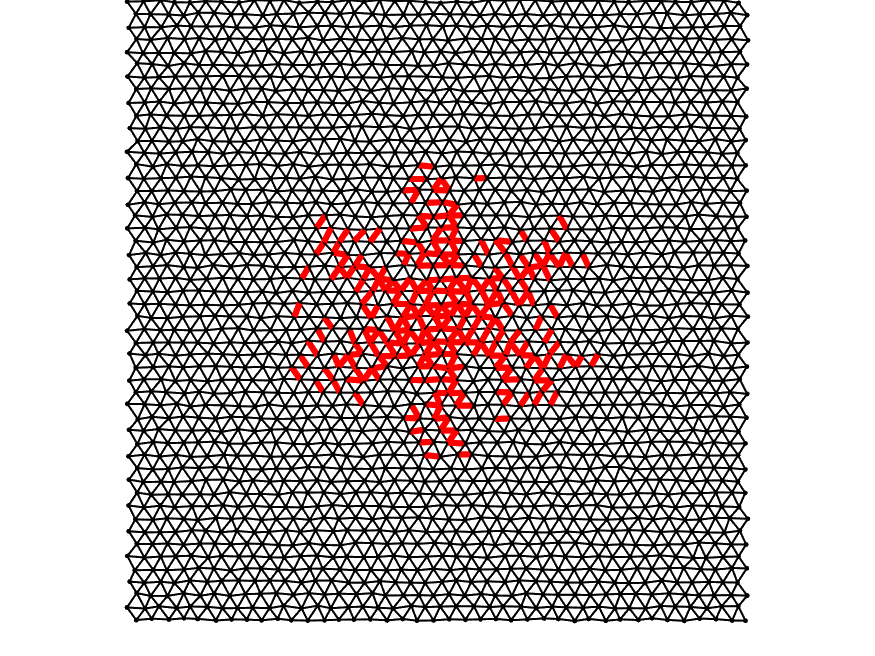}
    }
    \subfloat[]{
        \includegraphics[width=0.33\linewidth, clip=true, trim={40 0 60 0}]{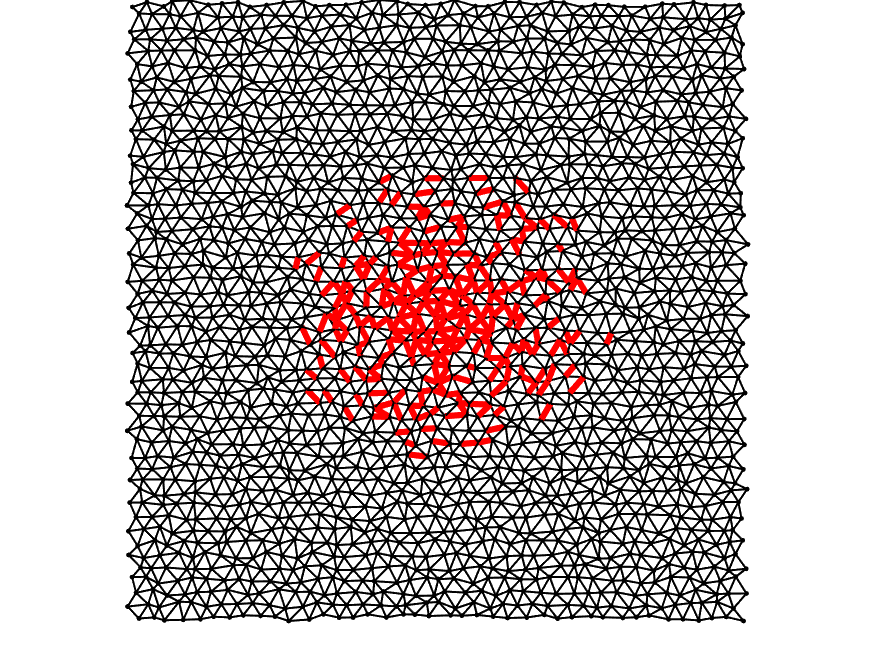}
    }
\caption{Lattice geometry and damage profile at a fixed time using the same plate area, quantity of material, energy of impact, link properties. Parameters are described in the main text. The broken links are plotted against the reference configuration in order to emphasize the damage pattern. The lattice nodes are perturbed by 1\%, 5\%, 10\% and 25\% of the reference link length according to a uniform random distribution. See discussion in Section \ref{section:randomization:pertnodes}. }%
\label{fig:rnd-mesh}
\end{figure}

Perhaps the simplest means of breaking translation symmetry of a periodic lattice is to perturb the location of the initial nodes. Thus we break the symmetry of the crack. This perturbation is achieved as follows: while keeping the available mass for constructing links fixed, we apply the transformation $x_{i} \to x_{i}(1 + \beta_{i})$, where the $\beta_{i}$ are each independent uniform random variables on the domain $[-q,q]$ for a fixed $q$. Some of the resulting damage profiles are shown in Figure \ref{fig:rnd-mesh}. 

The data collected from this experiment is shown in Table \ref{table:pertnodes} as independent variable $q$ is increased from $0$ to $0.25$, representing the largest allowed nodal perturbation. There, we see that larger variance in initial node positions increases energy dissipation while decreasing severity; for $q=0.236$, energy dissipation is maximized while severity is minimized. The radii associated with the 25th, 50th, 75th and 90th percentiles of energy dissipation are minimally affected. The ability to increase energy dissipation while decreasing damage severity strongly indicates the usefulness of perturbing node positions in improving resilience. As shown in Table \ref{table:summary}, this represents a simultaneous 17.5\% increase in energy dissipation and a 16.2\% decrease in severity of damage $S$ compared with the control lattice.

\begin{table}[htpb]
\centering
\begin{tabular}{||c c c c c c c c c c c||} 
 \hline
$q$ & 0	& 0.013 &	0.026 &	0.039 &	0.052 &	0.065 &	0.078 &	0.092 &	0.105 &	0.118 \\ [0.5ex] 
 \hline\hline
 $D$ & 0.097 & 0.095 & 0.097 & 0.096 & 0.0985 & 0.099 & 0.101 & 0.103 & 0.103 & 0.107\\ 
 $S$ & 0.376 & 0.363 & 0.345 & 0.333 & 0.333 & 0.333 & 0.328 & 0.322 & 0.320 & 0.320 \\
 $r_{.25}$ & \textbf{3.214} &	3.120 &	3.135 &	3.112 &	3.149 &	3.121 &	3.136 &	3.132 & 3.149 &	3.174 \\
 $r_{.50}$ & 5.131 &	4.967 &	4.985 &	4.961 &	4.951 &	4.985 &	4.987 &	4.997 &	4.999 &	5.055 \\
 $r_{.75}$ & \textbf{7.571} & \textit{7.513} &	7.322 &	7.187 &	7.184 &	7.111 &	7.081 &	7.177 &	7.147 &	7.206 \\
 $r_{.90}$ & \textbf{9.073} & \textit{8.975} &	8.968 &	8.719 &	8.754 &	8.711 &	8.613 &	8.715 & 8.607 &	8.695 \\ [1ex] 
 \hline
$q$ & 0.131 & 0.144 & 0.157 & 0.171 & 0.184 & 0.197 & 0.210 & 0.223 &	0.236 &	0.250 \\ [0.5ex] 
 \hline\hline
 $D$ & 0.107 & 0.108 & 0.109 & 0.108 & 0.112 & \textit{0.114} & 0.110 & 0.113 & \textbf{0.114} & 0.113 \\ 
 $S$ & 0.321 & 0.321 & 0.326 & 0.318 & 0.320 & 0.319 & 0.321 & \textit{0.317} & \textbf{0.315} & 0.324 \\
 $r_{.25}$ & 3.136 &	3.165 &	3.137 &	3.160 &	3.194 &	3.192 &	3.200 &	3.187 & \textit{3.206} &	3.190 \\
 $r_{.50}$ & 5.036 &	5.055 &	4.994 &	5.021 &	5.078 &	\textit{5.164} &	5.129 &	5.142 &	\textbf{5.196} &	5.139 \\
 $r_{.75}$ & 7.181 & 7.192 &	7.148 &	7.216 &	7.240 &	7.230 &	7.299 &	7.315 & 7.314 &	7.283 \\
 $r_{.90}$ & 8.578 & 8.701 &	8.603 &	8.669 &	8.698 &	8.712 &	8.738 &	8.880 & 8.840 &	8.744 \\ [1ex]  
 \hline\hline
\end{tabular}
\caption{ Summary of results for node perturbation experiment described in Section \ref{section:randomization:pertnodes}. The best values of $D, S, r_{.25}, r_{.50}, r_{.75}$ and $r_{.90}$ are \textbf{bolded}. The second best are \textit{italicized}. }
\label{table:pertnodes}
\end{table}

%\begin{figure}[htpb]
%\centering
%    \subfloat[]{
%        \includegraphics[width=0.33\linewidth, clip=true, trim={40 0 60 0}]{dissipation_pertnodes.eps}
%    }
%    \subfloat[]{
%     \includegraphics[width=0.33\linewidth, clip=true, trim={40 0 60 0}]{severity_pertnodes.eps}
%    } 
%    \subfloat[]{
%        \includegraphics[width=0.33\linewidth, clip=true, trim={40 0 60 0}]{radii_pertnodes.eps}
%    }
%    \caption{Results from randomly perturbing the default locations of links by $q\%$ according to discussion in Section \ref{section:pertnodes}. (a) Dissipation vs. perturbation magnitude $q$; (b) severity vs. perturbation magnitude $q$; (c) $r_{.25}, r_{.50}, r_{.75}$ and $r_{.90}$ vs. perturbation magnitude $q$. We find that randomly perturbing node locations induces a roughly 15\% increase in dissipated energy, 15\% decrease in severity, and has mixed effects on the spread of damage as measured by the energy dissipation radii. }%%
%\label{fig:pertnodes}
%\end{figure}

\subsection{Random perturbation of link strength} \label{section:randomization:pertk}
\begin{figure}[htpb]
\centering
    \subfloat[]{
        \includegraphics[width=0.33\linewidth, clip=true, trim={40 0 60 0}]{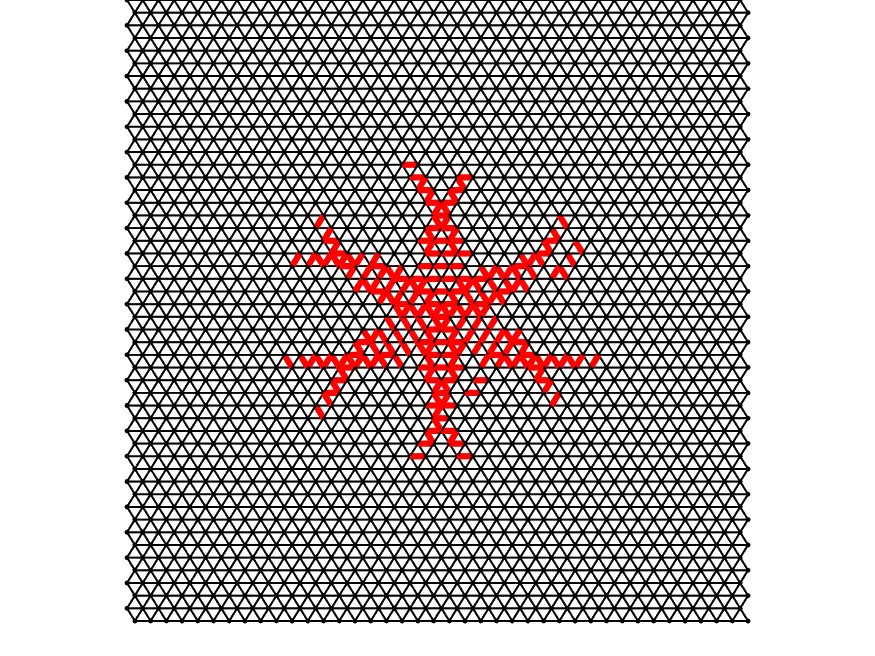}
    }
    \subfloat[]{
        \includegraphics[width=0.33\linewidth, clip=true, trim={40 0 60 0}]{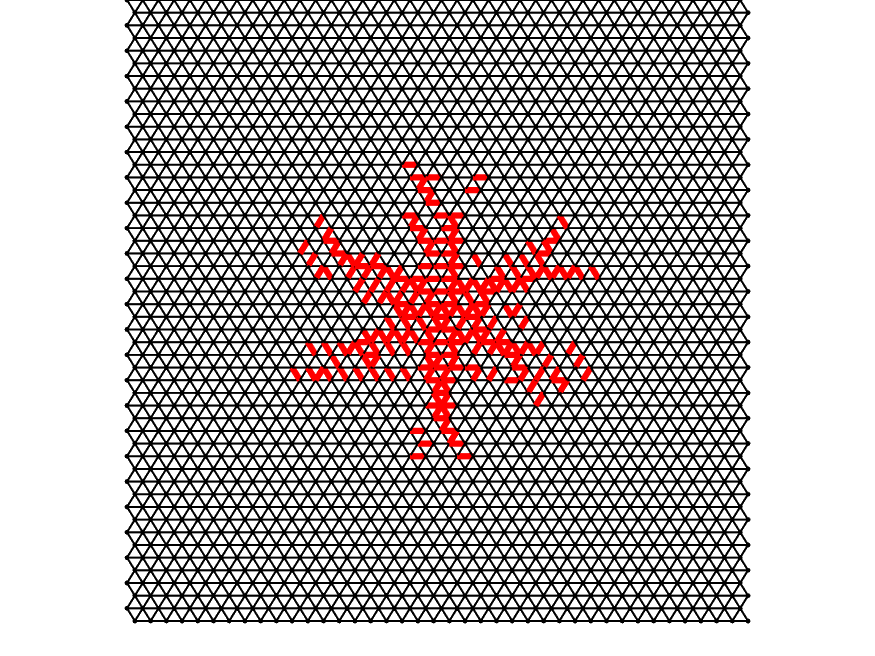}
    }
    \\
    \subfloat[]{
        \includegraphics[width=0.33\linewidth, clip=true, trim={40 0 60 0}]{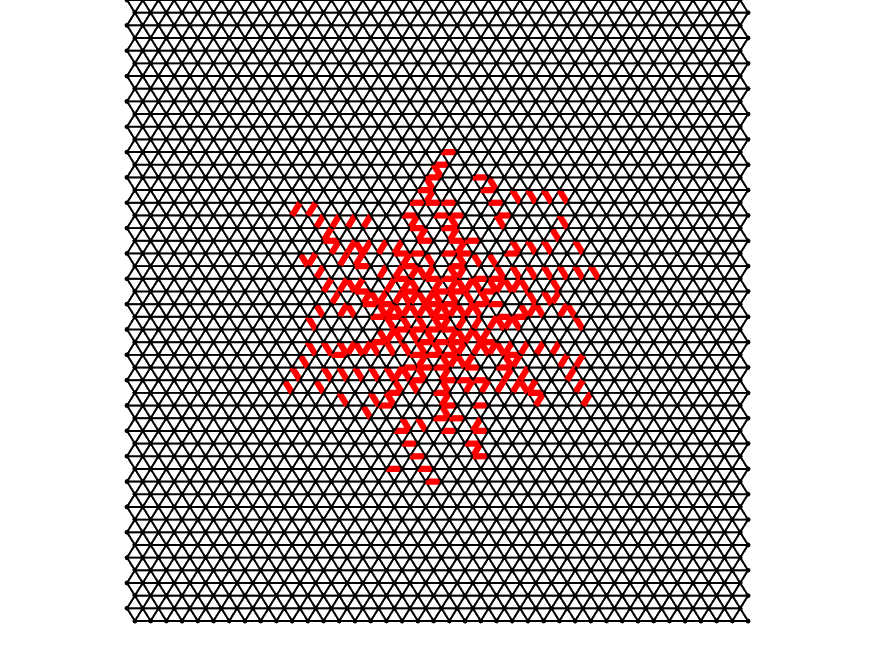}
    }
    \subfloat[]{
        \includegraphics[width=0.33\linewidth, clip=true, trim={40 0 60 0}]{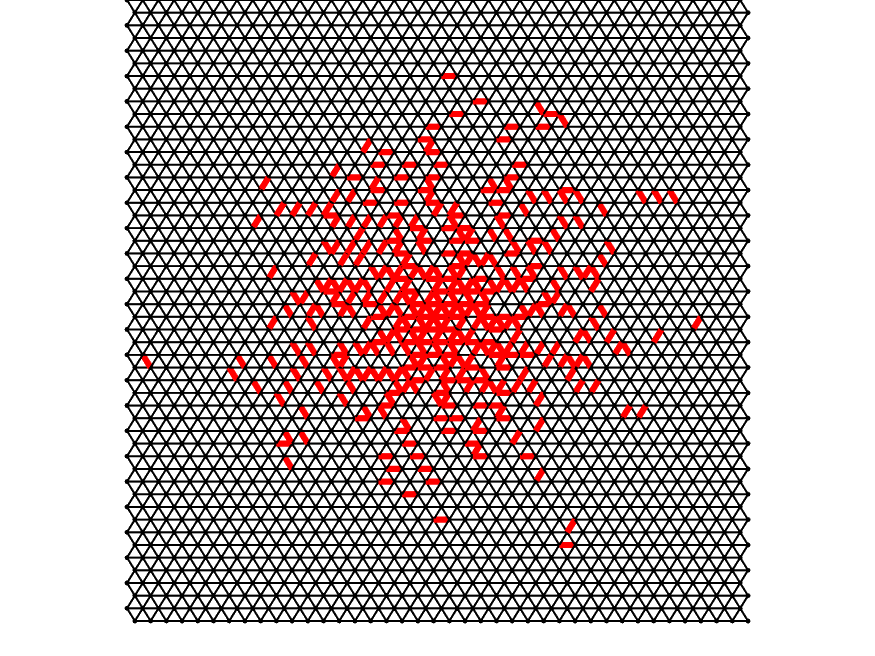}
    }
\caption{Lattice geometry and damage profile at a fixed time using the same plate area, quantity of material, energy of impact, link properties. Parameters are described in Section \ref{section:dynamics}. The broken links are plotted against the reference configuration in order to emphasize the damage pattern. Here, the stiffness $k$ of each link is randomly adjusted by up to (a) 1\%, (b) 10\%, (c) 50\%, (d) 100\% according to a uniform random distribution. See discussion in Section \ref{section:randomization:pertk}. }%
\label{fig:rnd-strength}
\end{figure}

Next, the strength the links is randomly perturbed according to a uniform random distribution while keeping the mean strength $k$ fixed. This is achieved by applying the transformation $k_{ij} \to k_{ij}(1 + \beta_{ij})$, where the $\beta_{ij}$ are each independent uniform random variables on the domain $[-q,q]$ for a fixed $q$. As $q$ is increased, the density of damage spreads over the domain. Some of the damage profiles resulting from this numerical experiment are shown in \Cref{fig:rnd-strength}. 

Data from this experiment is shown in Table \ref{table:pertk} as $q$ is adjusted from 0 to 1. When $q=1$, it is possible for links to have zero stiffness, or for links to double their original stiffness. There is a peak in energy dissipation at $q=0.421$, representing a modest 5\% increase in energy dissipation over the default construction. At $q=1$, there is a roughly 10\% decrease in energy dissipation compared to the default construction, while severity $S$ is minimized: a 20\% reduction compared with the default construction.

In contrast with the perturbation of node positions, perturbation of link strength leads to conflicting results: perturbing link strength increases $D$ modestly, but not as much as perturbing node positions, while it is possible to achieve much lower $S$, at the cost of dissipating less energy. The total damaged area, represented by the radii of dissipation $r_{p}$, is also generally increased for $q=1$. Hence perturbation of link strength can lead to less destruction of material and less concentrated damage. Whether this is desirable depends on the application.

\begin{table}[htpb]
\centering
\begin{tabular}{||c c c c c c c c c c c||} 
 \hline
$q$ & 0	& 0.052 & 0.105 & 0.157 & 0.210 & 0.263 & 0.315 & 0.368 & 0.421 & 0.473 \\ [0.5ex] 
 \hline\hline
 $D$ & 0.097 & 0.095 &	0.096 & 0.097 & 0.098 & 0.099 &	0.100 & \textit{0.100} & \textbf{0.101} & \textit{0.100} \\ 
 $S$ & 0.376 & 0.328 & 0.328 & 0.331 & 0.330 & 0.326 & 0.326 & 0.321 & 0.322 & 0.321 \\
 $r_{.25}$ & 3.214 &	3.392 &	3.242 &	3.214 &	3.274 &	3.328 &	3.388 &	3.388 &	3.388 &	3.447 \\
 $r_{.50}$ & 5.131 & 5.127 &	5.032 &	4.993 &	5.067 &	5.127 &	5.182 &	5.222 & 5.262 &	5.412 \\
 $r_{.75}$ & 7.571 &	6.991 &	6.820 &	6.658 &	6.688 &	6.804 &	6.832 &	6.930 &	7.018 &	7.060 \\
 $r_{.90}$ & 9.073 &	8.872 &	8.710 &	8.409 &	8.338 &	8.525 &	8.525 &	8.595 &	8.698 &	8.777 \\ [1ex] 
 \hline
$q$ & 0.526 & 0.578 & 0.631 & 0.684 & 0.736 & 0.789 & 0.842 & 0.894 & 0.947 & 1 \\ [0.5ex] 
 \hline\hline
 $D$ & 0.100 & 0.099 & 0.097 & 0.096 & 0.094 & 0.092 & 0.090 & 0.088 & 0.087 & 0.087 \\ 
 $S$ & 0.320 & 0.317 & 0.314 & 0.310 & 0.308 & 0.308 & 0.303 & 0.301 & \textit{0.301} & \textbf{0.299} \\
 $r_{.25}$ & 3.452 &	3.447 &	3.447 &	3.447 &	\textit{3.507} &	\textbf{3.562} &	3.502 &	\textit{3.507} & \textit{3.507} & \textit{3.507} \\
 $r_{.50}$ & 5.412 &	5.432 &	5.432 &	\textit{5.507} &	5.543 &	5.468 &	\textit{5.507} &	\textit{5.507} & \textbf{5.579} &	\textbf{5.579} \\
 $r_{.75}$ & 7.148 &	7.203 &	7.315 &	7.410 &	7.436 &	7.448 &	7.530 &	7.674 &	\textit{7.855} &	\textbf{7.931} \\
 $r_{.90}$ & 8.812 & 8.937 &	9.073 &	9.084 &	9.182 &	9.182 &	9.237 &	9.323 & \textit{9.472} &	\textbf{9.700} \\ [1ex]  
 \hline\hline
\end{tabular}
\caption{ Summary of results for link strength perturbation experiment described in Section \ref{section:randomization:pertk}. The best values of $D, S, r_{.25}, r_{.50}, r_{.75}$ and $r_{.90}$ are \textbf{bolded}. The second best are \textit{italicized}.  }
\label{table:pertk}
\end{table}

%\begin{figure}[htpb]
%\centering
%    \subfloat[]{
%        \includegraphics[width=0.33\linewidth, clip=true, trim={40 0 60 0}]{dissipation_pertk.eps}
%    }
%    \subfloat[]{
%     \includegraphics[width=0.33\linewidth, clip=true, trim={40 0 60 0}]{severity_pertk.eps}
%    } 
%    \subfloat[]{
%        \includegraphics[width=0.33\linewidth, clip=true, trim={40 0 60 0}]{radii_pertk.eps}
%    }
%    \caption{Results from randomly perturbing stiffnesses of links according to discussion in Section \ref{section:pertk}. (a) Dissipation vs. perturbation magnitude $q$; (b) severity vs. perturbation magnitude $q$; (c) $r_{.25}, r_{.50}, r_{.75}$ and $r_{.90}$ vs. perturbation magnitude $q$. We find that randomly perturbing link strength induces a 5\% increase in dissipated energy at peak and a 20\% decrease in severity. The effects on radii of dissipation are mixed: around a 10\% largest allowed perturbation, the radii are at their smallest.}%%
%\label{fig:pertk}
%\end{figure}

\subsection{Random removal of links} \label{section:randomization:removal}
\begin{figure}[htpb]
\centering
    \subfloat[]{
        \includegraphics[width=0.33\linewidth, clip=true, trim={40 0 60 0}]{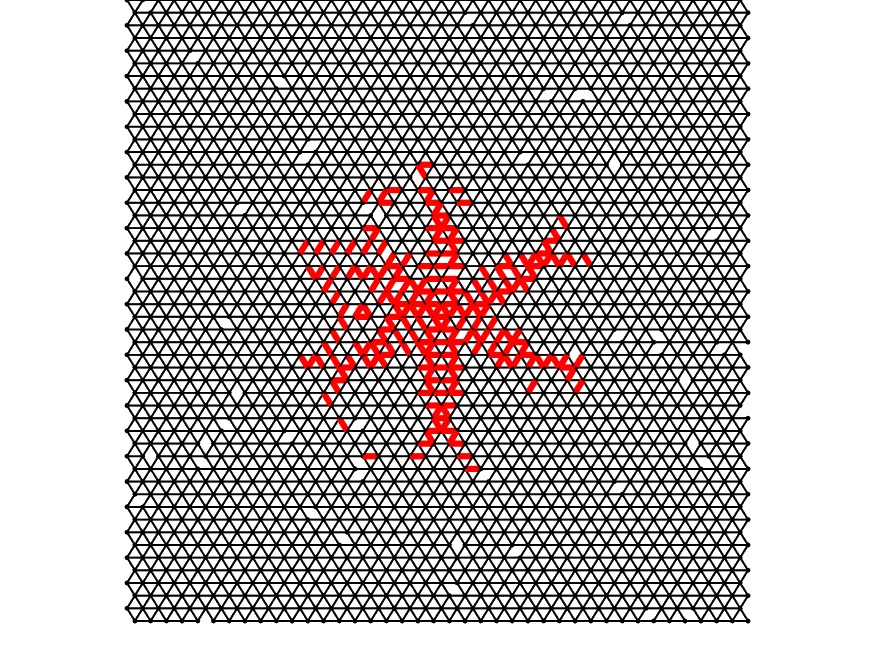}
    }
    \subfloat[]{
     \includegraphics[width=0.33\linewidth, clip=true, trim={40 0 60 0}]{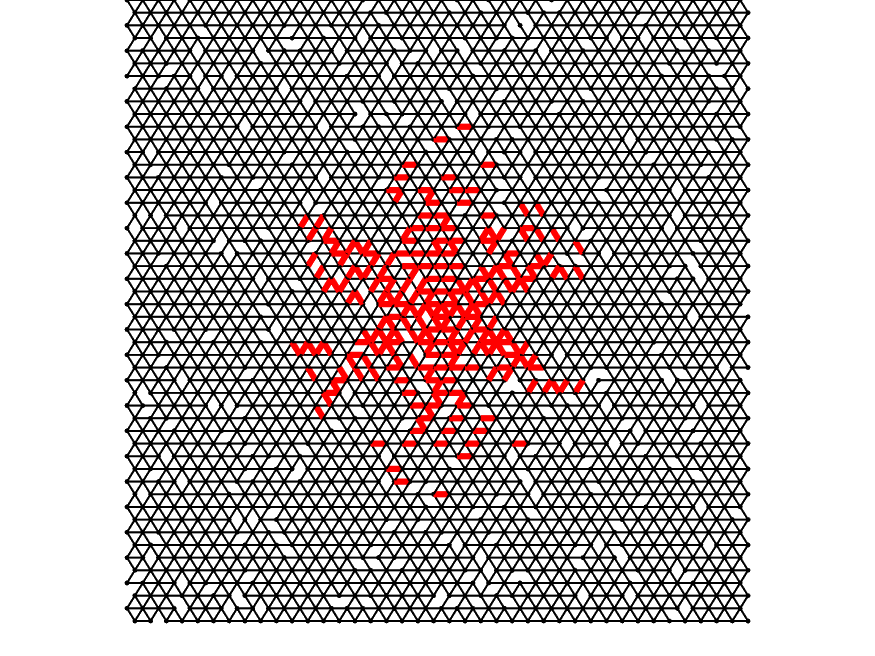}
    }
    \\
    \subfloat[]{
     \includegraphics[width=0.33\linewidth, clip=true, trim={40 0 60 0}]{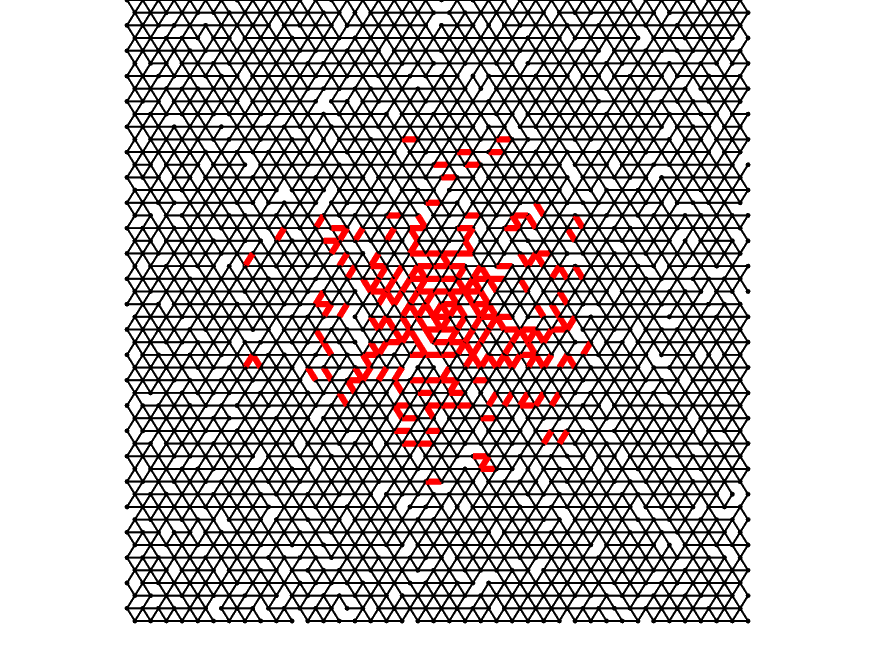}
    }
    \subfloat[]{
     \includegraphics[width=0.33\linewidth, clip=true, trim={40 0 60 0}]{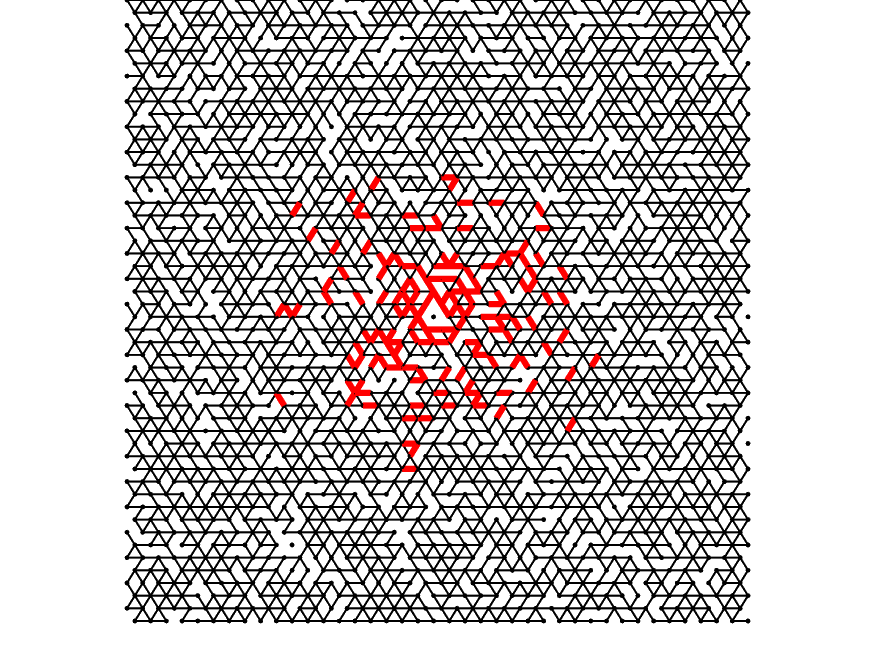}
    }
    \caption{Lattice geometry and damage profile at a fixed time using the same plate area, quantity of material, energy of impact, link properties. Parameters are described in Section \ref{section:dynamics}. The broken links are plotted against the reference configuration in order to emphasize the damage pattern. Here, links are removed from the initial construction according to a uniform random variable and the mass of those links is reallocated throughout the remaining structure. (a) 1\% removal; (b) 5\% removal; (c) 10\% removal; (d) 25\% removal. See discussion in Section \ref{section:randomization:removal}. }%
\label{fig:removal}
\end{figure}

Next, we consider randomly removing a fraction of links according to a uniform random distribution, seeking to break symmetry and deny the propagating crack a natural direction of propagation. This is done by assigning realizations of a uniform random variable on the domain $[0,1]$ to each link $(ij)$. If the value of a given realization is below a fixed and predetermined threshold $q$, the link $(ij)$ is deleted. Hence $q$ is the fraction of removed links. In order to conserve the amount of material used, the mass of removed links is uniformly reallocated to the rest of the construction. The results of these experiments are shown visually in Figure \ref{fig:removal}. The collected data is shown in Table \ref{table:removal} as threshold is adjusted $q$ from 0 to 0.25, representing an increase in the removed fraction from 0\% to 25\%. 

There is a peak in energy dissipation at 7.8\% removal, representing a 12.3\% increase above the baseline construction. The severity $S$ tends to decrease with $q$; a 20.4\% decrease is also observed at $q=0.078$. Trends in the radii associated with the 25th, 50th, 75th and 90th percentiles are mixed, with the 25th and 50th percentile radii reflecting a modest expansion of the region over which energy is dissipated. These results recall the use of holes and other inhomogeneities to modify crack development in elastic media \cite{xu-etal-MSMSE-1997,mogilevskaya-et-al-IJOF-2009,huang-et-al-PMS-2015,damaskinskaya-et-al-POSS-2018}. Random link removal increases dissipated energy, but not as much as node perturbation, and decreases severity $S$, but not as much link strength perturbation. However, random removal achieves this reduced severity without decreasing dissipated energy, unlike link strength perturbation.

\begin{table}[htpb]
\centering
\begin{tabular}{||c c c c c c c c c c c||} 
 \hline
$q$ & 0	& 0.013 &	0.026 &	0.039 &	0.052 &	0.065 &	0.078 &	0.092 &	0.105 &	0.118 \\ [0.5ex] 
 \hline\hline
 $D$ & 0.097 & 0.098 & 0.102 & 0.104 & 0.106 & 0.104 & \textbf{0.109} & 0.105 & 0.106 & \textit{0.107} \\ 
 $S$ & 0.376 & 0.335 & 0.318 & 0.316 & 0.307 & 0.299 & 0.304 & 0.300 & 0.292 & 0.289 \\
 $r_{.25}$ & 3.214 & 3.363 &	3.477 &	3.564 &	\textbf{3.619} &	3.564 &	3.562 &	3.532 & 3.557 &	\textit{3.587} \\
 $r_{.50}$ & 5.131 & 5.164 &	5.240 &	\textbf{5.450} & \textbf{5.450} & 5.402 &	\textit{5.448} & 5.277 & 5.410 & 5.440 \\
 $r_{.75}$ & \textbf{7.571} & 7.002 &	7.180 &	7.284 &	7.257 &	7.230 &	7.254 &	7.100 &	7.178 &	7.244 \\
 $r_{.90}$ & \textit{9.073} &	8.497 &	8.618 &	8.796 &	8.853 &	8.743 &	8.795 &	8.743 &	8.762 &	8.734 \\ [1ex] 
 \hline
$q$ & 0.131 & 0.144 & 0.157 & 0.171 & 0.184 & 0.197 & 0.210 & 0.223 &	0.236 &	0.250 \\ [0.5ex] 
 \hline\hline
 $D$ & 0.102 & 0.101 & 0.102 & 0.096 & 0.094 & 0.088 & 0.089 & 0.090 & 0.084 & 0.083 \\ 
 $S$ & 0.277 & 0.282 & 0.280 & 0.273 & 0.276 & 0.267 & \textbf{0.261} & 0.268 & \textit{0.264} & \textit{0.264} \\
 $r_{.25}$ & 3.534 &	3.500 &	3.534 &	3.317 &	3.358 &	3.164 &	3.342 &	3.239 & 3.162 &	3.086 \\
 $r_{.50}$ & 5.418 &	5.401 &	5.412 &	5.248 &	5.239 &	5.010 &	5.250 &	4.955 & 4.935 &	4.742 \\
 $r_{.75}$ & \textit{7.413} &	7.349 &	7.243 &	7.145 &	6.950 &	6.890 &	6.916 &	6.859 &	6.880 &	6.567 \\
 $r_{.90}$ & \textbf{9.085} &	8.819 &	8.997 &	8.688 &	8.729 &	8.557 &	8.614 &	8.663 &	8.415 &	8.382 \\ [1ex]  
 \hline\hline
\end{tabular}
\caption{ Summary of results for random link removal experiment described in Section \ref{section:randomization:removal}. The best values of $D, S, r_{.25}, r_{.50}, r_{.75}$ and $r_{.90}$ are \textbf{bolded}. The second best are \textit{italicized}.  }
\label{table:removal}
\end{table}

%\begin{figure}[htpb]
%\centering
%    \subfloat[]{
%        \includegraphics[width=0.33\linewidth, clip=true, trim={40 0 60 0}]{dissipation_removal.eps}
%    }
%    \subfloat[]{
%     \includegraphics[width=0.33\linewidth, clip=true, trim={40 0 60 0}]{severity_removal.eps}
%    } 
%    \subfloat[]{
%        \includegraphics[width=0.33\linewidth, clip=true, trim={40 0 60 0}]{radii_removal.eps}
%    }
%    \caption{Results from randomly removing $q$\% of links according to discussion in Section \ref{section:removal}. (a) Dissipation vs. removal rate $q$; (b) severity vs. removal rate $q$; (c) the 25th, 50th, 75th and 90th percentile radii of dissipation vs. removal rate $q$. We find that random removal of links produces a peak 10\% increase in energy dissipation when around 8\% of links are removed and have their mass reallocated. Severity decreases by around 20\% from baseline around this same value. Effects on radii of dissipation are mixed. }%%
%\label{fig:removal2}
%\end{figure}

\subsection{Combined randomization}
\label{section:randomization:combinationofeffects}
Until now, the effects of stiffness perturbation, node position perturbation, and random removal have been considered individually. In order to investigate whether it is possible to combine these effects to achieve results superior to a single design intervention alone, or whether they counteract each other, we consider simulations of a lattice with (a) an 8\% removal rate, (b) perturbation of stiffness chosen as a uniform random variable with $q=.5$, and (c) perturbation of node positions with $q=.25$. These values of $q$ each individually maximized energy dissipation above. The results of this experiment are as follows.
\begin{itemize}
    \item 10.54\% of impact energy is dissipated, an improvement on stiffness perturbation alone, while not as good as perturbation of node positions alone. 
    \item A severity of 0.29, which is lower than that of any of stiffness perturbation, location perturbation, and random removal individually at the values of $q$ used in this experiment. Hence there are further reductions in severity by combining different kinds of randomization.
    \item 25th, 50th, 75th and 90th percentile radii of energy dissipation as 3.16, 5.40, 7.60 and 9.70 respectively. These values represent a significant expansion of the area over which energy is dissipated compared to baseline, and are also greater than those found for each of the design interventions considered individually.
\end{itemize}
These results show that it may be possible to ``hybridize" the approaches discussed in this paper in order to achieve even better performance. See also Table \ref{table:summary}.

\begin{figure}[htpb]
\centering
    \subfloat[]{
        \includegraphics[width=0.33\linewidth, clip=true, trim={40 0 60 0}]{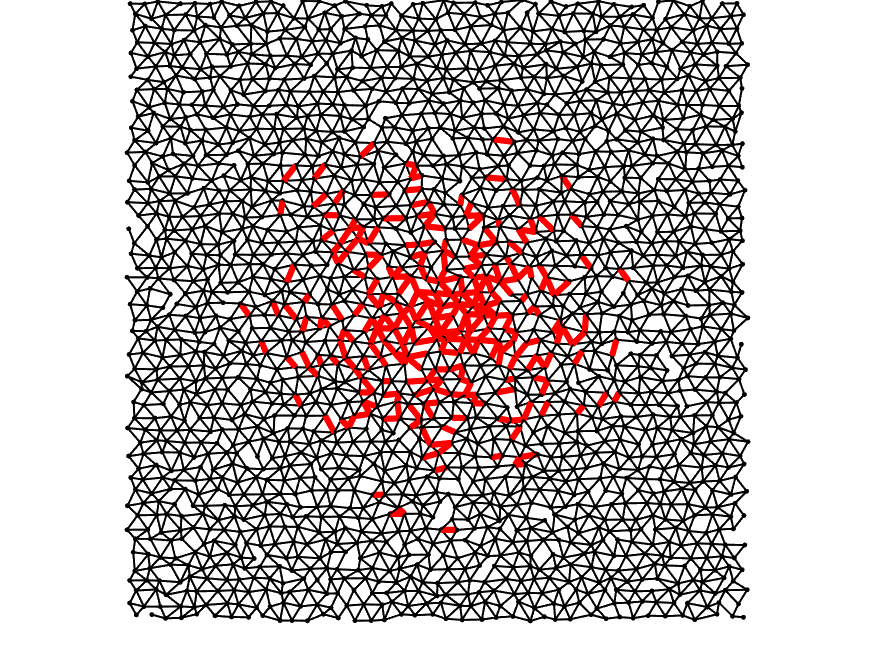}
    }
    \subfloat[]{
        \includegraphics[width=0.33\linewidth, clip=true, trim={40 0 60 0}]{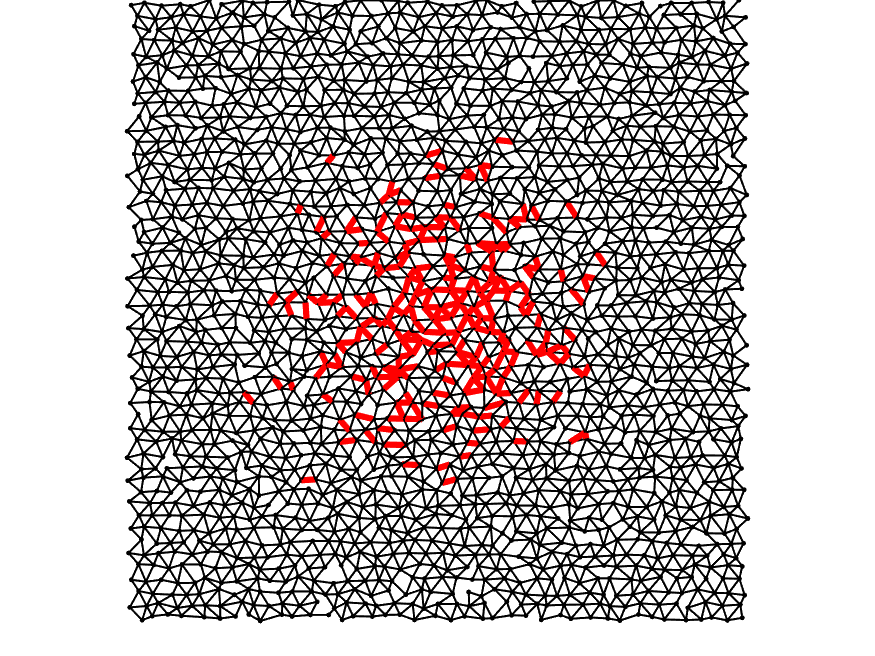}
    }
    \caption{The resulting damage profile for two realizations of a lattice randomized according to discussion in Section \ref{section:randomization:combinationofeffects}. The random removal rate, perturbations of $k$ and perturbations of node positions are applied at their optimizing values from within each of Sections \ref{section:randomization:pertnodes}, \ref{section:randomization:pertk} and \ref{section:randomization:removal}.  }%%
\label{fig:optimalish}
\end{figure}

\section{Bistable links} \label{section:sacrificial}
\subsection{Initial results}
Here, we consider an additional means of symmetry-breaking in lattices whose links contain sacrificial elements--- components intended to be broken in order to improve overall structural robustness \cite{slepyan-ayzenberg-stepanenko-JMPS-2004,cherkaev-et-al-JMMPS-2005,cherkaev-at-al-MOM-2006,cherkaev-leelavanichkul-IJODM-2012,nadkarni-et-al-PRE-2014}. Breakage of a sacrificial link causes a local instability, which has two immediate effects: first, it dissipates energy without destruction of the link, and, second, it redistributes stress to neighboring links. Physically, each link consists of two almost-parallel elasto-brittle rods. One is slightly curved and begins to resist after the elongation reaches a threshold. Mathematically, the force-elongation dependence is 
\begin{equation}
s(e,t)=\left\{
\begin{array}{cl}
k_se  &    t < t_s' \cr
k_w(e-e_w) &  t_s' \leq t < t_w' \cr
0 & t \geq t_w'
\end{array}
\right.
\end{equation}
where $t_s'$ is the first time at which $e>e_w$ (i.e., the time of breakage for the sacrificial link), and $t_w'$ is the first time at which $e>e_{\text{max}}$ (i.e., the time of breakage of the waiting link). $e_w$ is the activation threshold at which the sacrificial link breaks and the waiting link instantaneously takes over the load. Links are completely destroyed at 20\% elongation, hence $e_{\text{max}}$ = 0.2.

The total energy dissipated by breaking both the sacrificial and waiting components of the link is fixed. It is required that
\begin{equation}
    \int_{0}^{e_{\text{max}}}s(e)de = E_{bi},
\end{equation}
where $E_{bi}$ is specified. The above model then depends on only two parameters: (i) the fraction $\alpha$ of material placed in the first (sacrificial) link and (ii) the size $e_w$ of the threshold. Selecting $\alpha$ by allocating mass to the first (sacrificial) link fixes $k_s$. Then, fixing the threshold $e_w$ fully determines $k_w$. Since (a) the stiffness of a link is proportional to its cross-sectional area with proportionality constant 10, (b) the waiting link activates at elongation $e_w$, and (c) a fixed amount of mass $m$ is available to allocate to each compound link, $k_s = 10\frac{m_s}{L_s}$ and $k_w = 10\frac{m_w}{L_w}$
where $m_s + m_w = m$. $L_s=1$ and $L_w = 1+e_w$. As with the previous numerical experiments described in this paper, the sacrificial link is of unit length. The waiting link activates at $e_w$ elongation beyond this, hence $L_s + L_w = 2 + e_w$. 

As before, a 40-by-40 triangular lattice is constructed which satisfies these specifications, taking lattice mass $M=10000$ units. It is assumed that mass is distributed evenly among the compound links (i.e., the combination of the sacrificial and waiting link), resulting in mass $m$ available for each such compound link. Acceptable parameter values $\alpha, e_w$ lie on the curve
\begin{equation}
    10\big(\alpha m + \frac{(1-\alpha)m}{1+e_w}\big)e_w^2 + 10\frac{(1-\alpha)m}{1+e_w}\big(\frac{e_{max}^2}{2} - e_{max}e_w\big) = E_{bi}, \label{curve}  
\end{equation}
where $0 < \alpha < 1$, and $0 < e_w < e_{max}$.

\begin{figure}[htpb]
\centering
    \subfloat[]{
        \includegraphics[width=0.33\linewidth, clip=true, trim={40 0 60 0}]{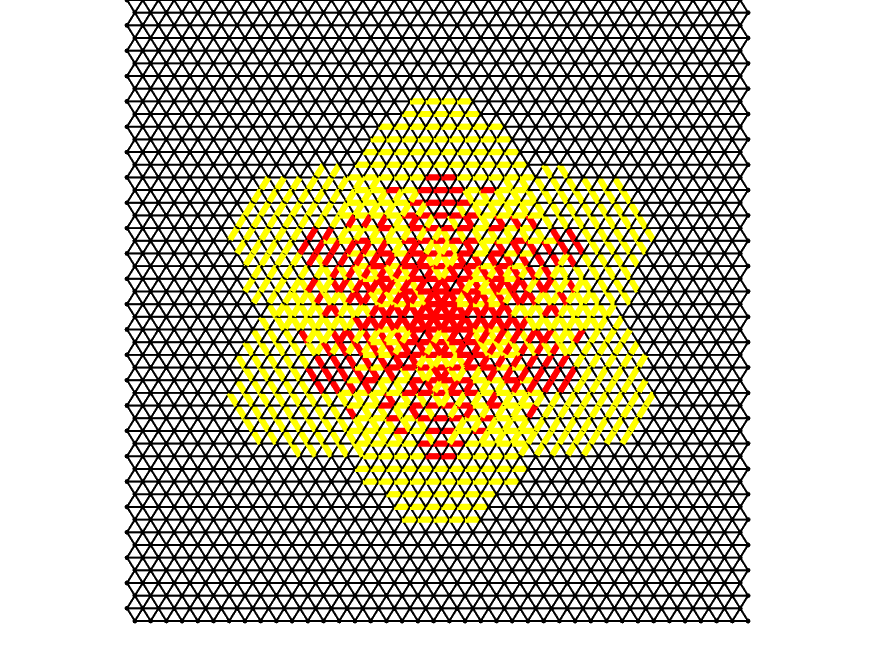}
    }
    \subfloat[]{
     \includegraphics[width=0.33\linewidth, clip=true, trim={40 0 60 0}]{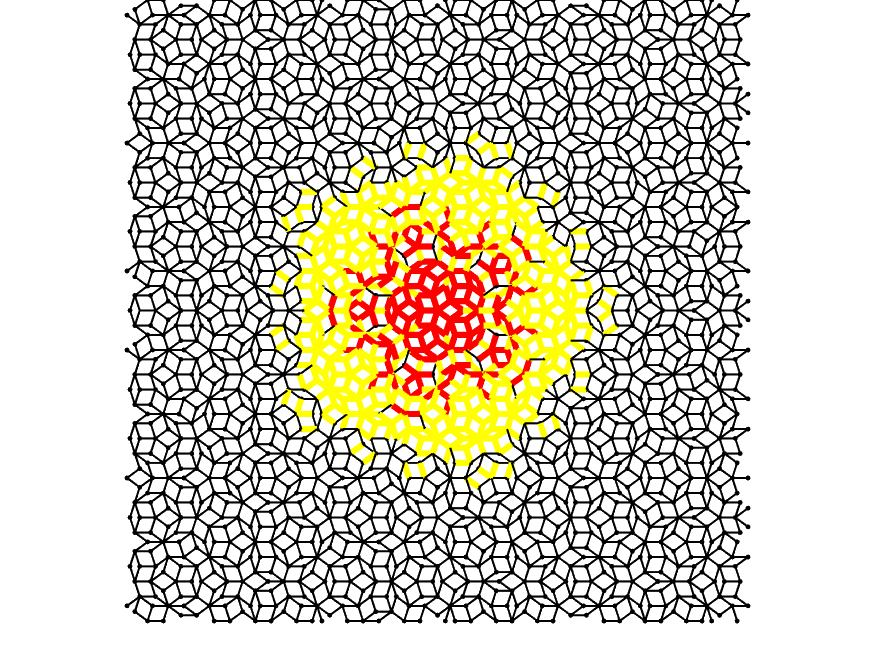}
    }
    \caption{Bistable triangular lattice and bistable Penrose lattice with $\alpha=0.25$ and $e_w=0.0547$. These parameters are chosen for aesthetic value, and to emphasize the delocalization of damage induced by the waiting link structure. The total energy dissipated by entirely destroying a link is 0.6502 units. Links colored red are completely destroyed, as elsewhere in the text; links colored yellow are damaged, with the sacrificial link broken but the waiting link surviving. See discussion in Section \ref{section:sacrificial}. }%
\label{fig:bistable1}
\end{figure}

\begin{figure}[htpb]
\centering
    \subfloat[]{
        \includegraphics[width=0.33\linewidth, clip=true, trim={40 0 60 0}]{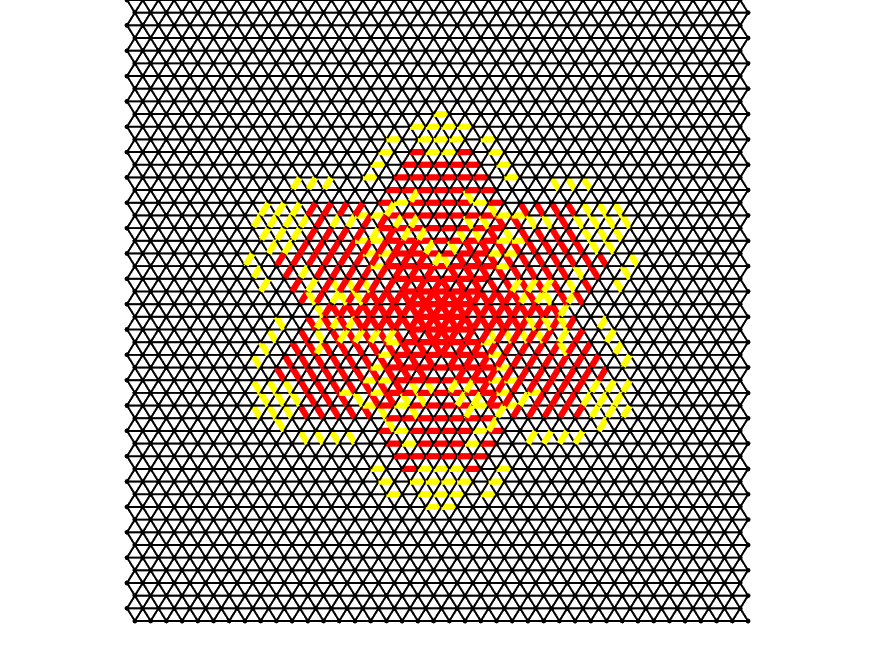}
    }
    \subfloat[]{
        \includegraphics[width=0.33\linewidth, clip=true, trim={40 0 60 0}]{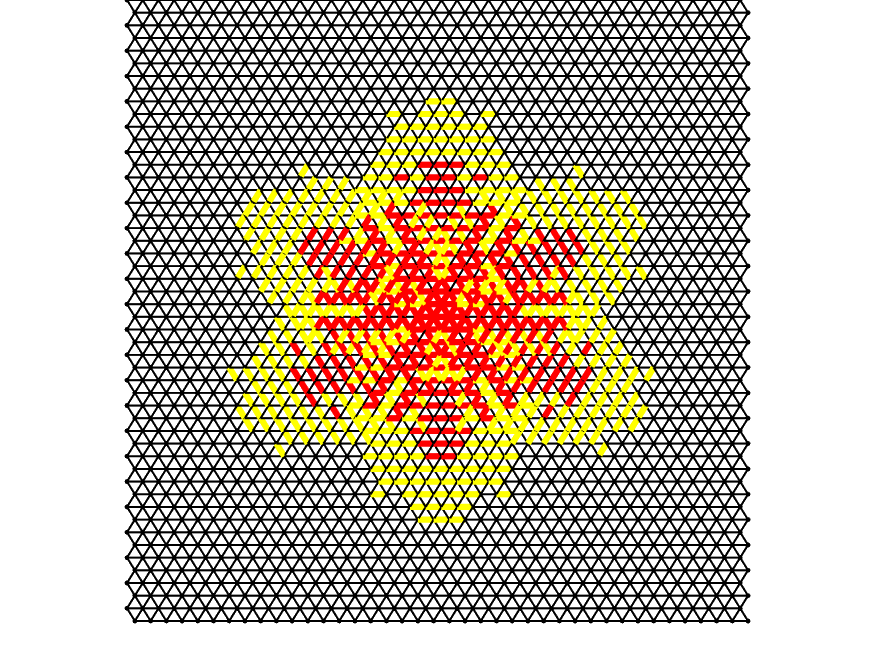}
    } \\ 
    \subfloat[]{
        \includegraphics[width=0.33\linewidth, clip=true, trim={40 0 60 0}]{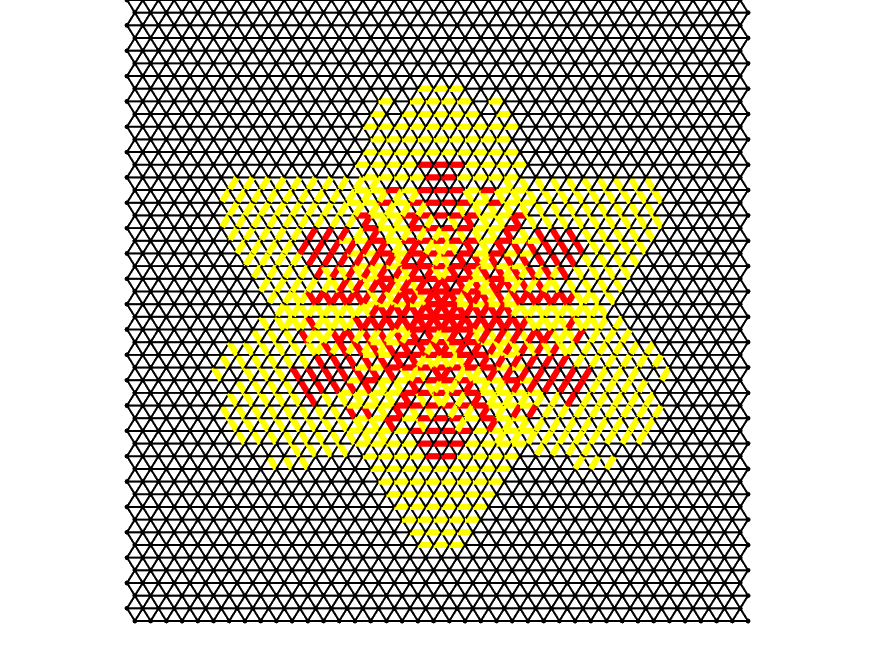}
    } 
    \subfloat[]{
        \includegraphics[width=0.33\linewidth, clip=true, trim={40 0 60 0}]{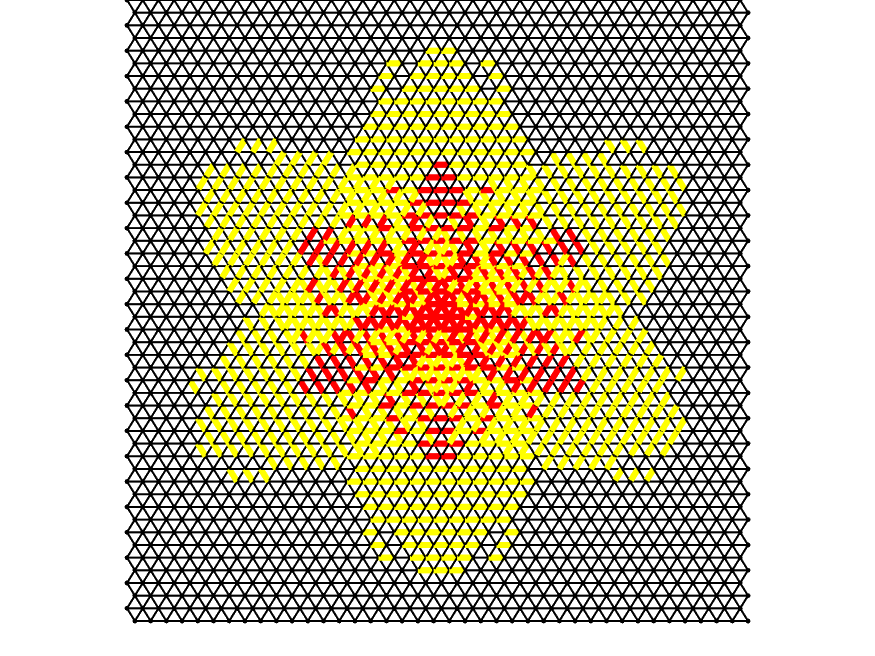}
    }
    \caption{The damage profiles resulting from an initial impulse of 800 units of energy. The total energy release by the destruction of a link with waiting structure is 0.1899 units with different thresholds for waiting link activation $e_w$ and mass fraction $\alpha$ allocated to the sacrificial link: (a) $e_w = 0.08, \alpha = 0.45$; (b) $e_w = 0.06, \alpha = 0.3765$; (c) $e_w = 0.04, \alpha = 0.427$; (d) $e_w = 0.02, \alpha = 0.504$. Links colored red are destroyed, as elsewhere in the text; links colored yellow are damaged, with the sacrificial link broken but the waiting link surviving. See discussion in Section \ref{section:sacrificial}. }%%
\label{fig:bistable2}
\end{figure}

Results for a triangular lattice and Penrose lattice are shown in Figure \ref{fig:bistable1}, where the energy dissipated by the complete destruction of a link (both the sacrificial and waiting components) is $E_{bi}=0.6502$. This value is chosen arbitrarily. It can be seen that the inclusion of the bistable links leads to an improved damage profile: the severity of the destroyed cluster is reduced, and many links are damaged but not destroyed. The same general tendency is seen in Figure \ref{fig:bistable2} for a bistable triangular lattice with $E_{bi} = 0.1899$. Due to the form of (\ref{curve}), many different values of $e_w$ and $\alpha$ are possible for the same $m,e_{max}$ and $E_{bi}$. Although the optimization of $e_w$ and $\alpha$ is not discussed here, it is evident that the cluster of destroyed links in Figure \ref{fig:bistable2}(a) is larger than that of Figure \ref{fig:bistable2}(d), suggesting that optimal values may be chosen.

\subsection{Perturbation of sacrificial and waiting link strength} \label{section:pertk-bistable}
\begin{figure}[htpb]
\centering
    \subfloat[]{
        \includegraphics[width=0.33\linewidth, clip=true, trim={40 0 60 0}]{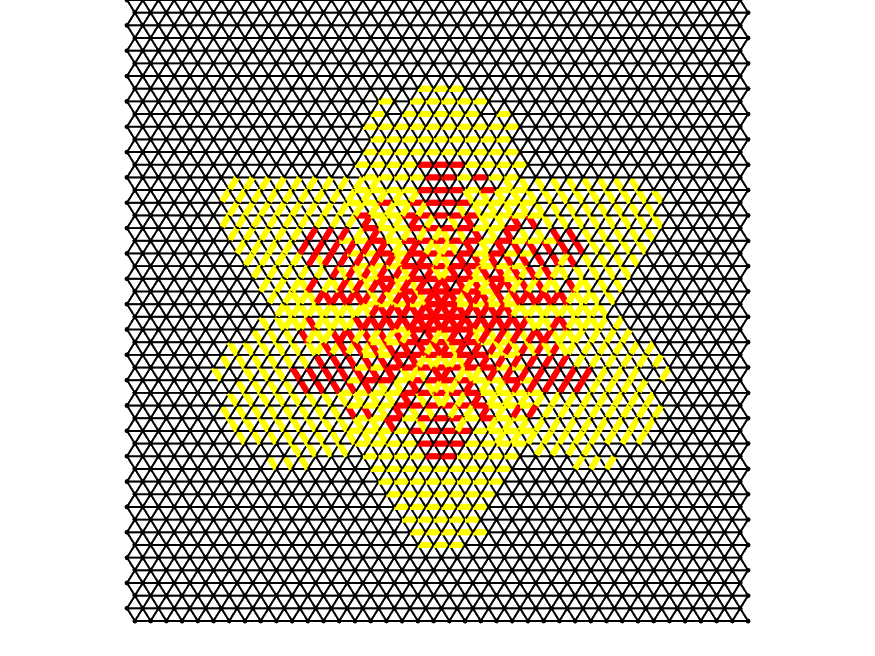}
    }
    \subfloat[]{
     \includegraphics[width=0.33\linewidth, clip=true, trim={40 0 60 0}]{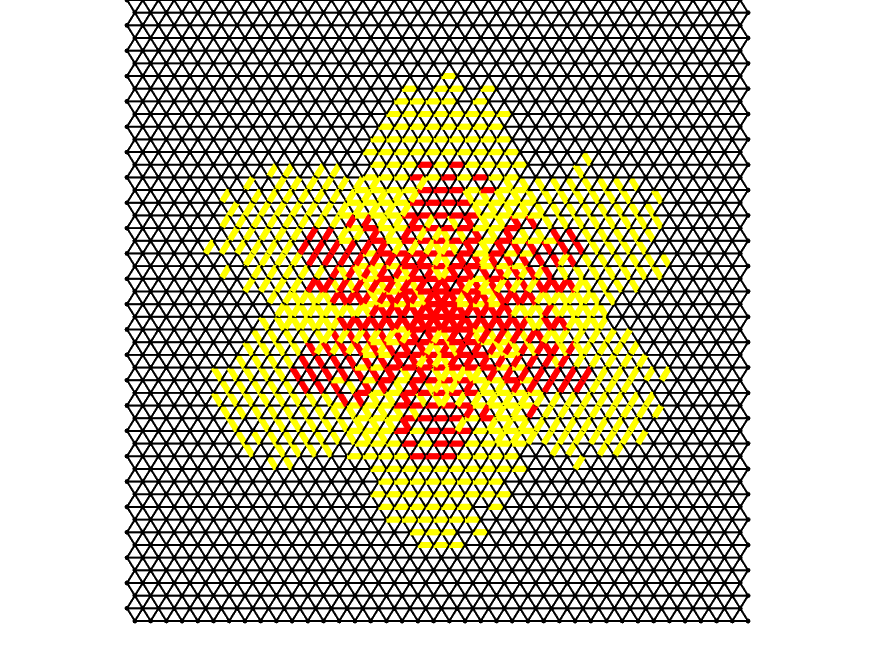}
    } \\
    \subfloat[]{
        \includegraphics[width=0.33\linewidth, clip=true, trim={40 0 60 0}]{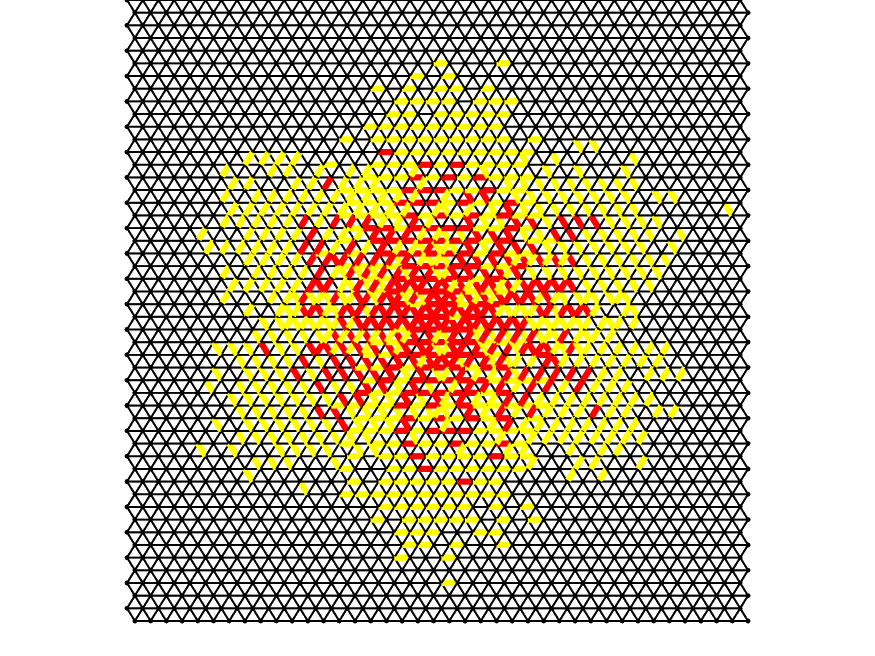}
    }
    \subfloat[]{
     \includegraphics[width=0.33\linewidth, clip=true, trim={40 0 60 0}]{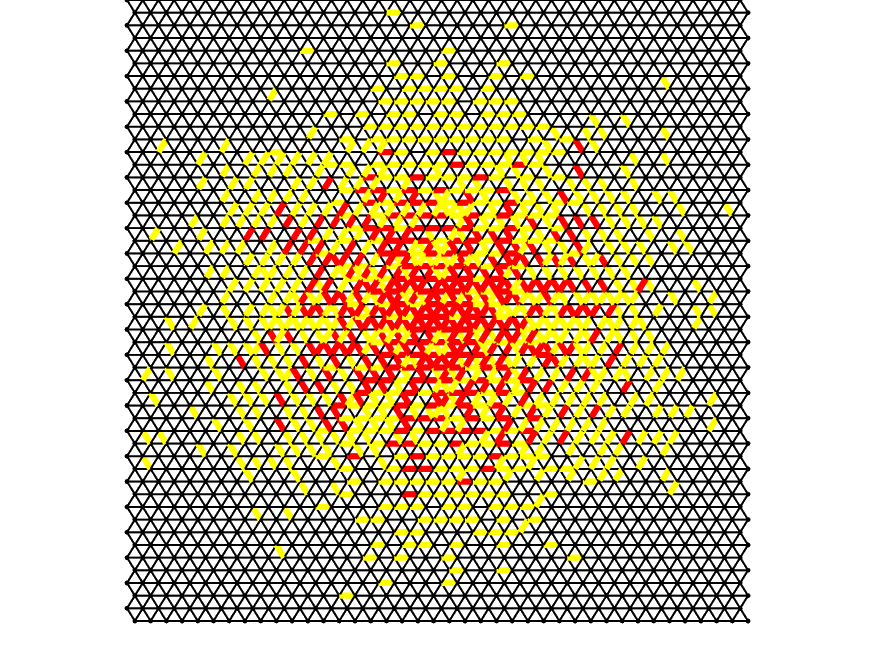}
    }
    \caption{Damage profile of bistable links with average $e_w=0.04, \alpha=0.427$. The stiffness $k_s$ and $k_w$ of individual components are perturbed by up to (a) 1\%, (b) 10\%, (c) 50\%, and (d) 100\%. Links colored red are destroyed, as elsewhere in the text; links colored yellow are damaged, with the sacrificial link broken but the waiting link surviving. See discussion in Section \ref{section:pertk-bistable}. }%%
\label{fig:bistable3}
\end{figure}

Keeping with the line of inquiry established previously for elasto-brittle links, we briefly consider perturbation of the stiffnesses $k_s$ and $k_w$ for the individual components of the waiting link structure. Some of these simulation results are presented in Figure \ref{fig:bistable3} for the values $e_w = 0.04$ and $\alpha = 0.427$. It can be seen that increasing the magnitude of the perturbation of $k_s$ and $k_w$ leads to an increase in the damaged area and a reduction in connectivity of the destroyed cluster near the impact, hence a reduction in severity $S$. This illustrates, broadly, that the symmetry-breaking principle discussed above applies also to bistable links.

Table \ref{table:pertk-bistable} shows collected data on energy dissipation, severity of damage and the radii of dissipation as the stiffnesses $k_w$ and $k_s$ are perturbed. $E_{bi} = 0.4645$, the same energy loss per destroyed link as in Sections \ref{section:randomization:pertnodes} - \ref{section:randomization:combinationofeffects}, so that the effects of the waiting link structure and randomization are isolated.

Unlike the simple elasto-brittle links studied in the rest of this paper, here the perturbation of stiffness leads to reductions in energy dissipation. However, the severity of damage is also reduced as the magnitude of perturbation increases; hence it is possible to randomize bistable lattices and exchange energy dissipation for reductions in the size of the broken cluster. Effects on the radii of dissipation are, again, mixed but minor. 

Note that, although the energy dissipation is reduced by perturbing $k_w$ and $k_s$, the energy dissipation at the largest perturbation strength is 16\%--- around 60\% higher than for elasto-brittle lattice without waiting structure. At the same time, the severity $S$ is comparable to that achieved by the optimal perturbation of $k$ for the elasto-brittle lattice without waiting structure: about .32 without waiting structure, and .33 with waiting structure. This shows that even when controlling for lattice mass $M$ and energy lost per link $E_{\text{loss}}$, it is possible for a bistable lattice to dissipate 50\% more energy with radii of dissipation and severity comparable to optimally-randomized lattices without waiting links. Because a single criterion to characterize resilience does not yet exist, it is unclear to what extent energy dissipation should be exchanged in favor of reduced severity of damage; however, Table \ref{table:pertk-bistable} shows that randomization of link strength is a means of tuning this exchange in bistable lattices.

\begin{table}[htpb]
\centering
\begin{tabular}{||c c c c c c c c c c c||} 
 \hline
$q$ & 0	& 0.052 & 0.105 & 0.157 & 0.210 & 0.263 & 0.315 & 0.368 & 0.421 & 0.473 \\ [0.5ex] 
 \hline\hline
 $D$ & \textbf{0.235} & \textit{0.234} & 0.232 & 0.230 & 0.228 & 0.225 & 0.222 & 0.218 & 0.215 & 0.212 \\ 
 $S$ & 0.384 & 0.376 & 0.376 & 0.372 & 0.367 & 0.365 & 0.363 & 0.360 & 0.357 & 0.353 \\
 $r_{.25}$ & \textbf{4.041} & \textbf{4.041} &	\textbf{4.041} &	\textbf{4.041} &	\textbf{4.041 }&	\textbf{4.041} &	\textbf{4.041} &	\textbf{4.041} & \textbf{4.041} &	\textbf{4.041 }\\
 $r_{.50}$ & \textbf{5.859} &	\textit{5.842} &	\textbf{5.859} &	5.825 &	5.807 &	5.790 &	5.790 &	5.773 & 5.738 &	5.721 \\
 $r_{.75}$ & \textbf{8.082} &	7.919 &	7.715 &	7.650 &	7.637 &	7.585 &	7.637 &	7.676 & 7.702 &	7.741 \\
 $r_{.90}$ & 9.291 & 9.269 &	9.139 &	9.073 &	9.062 &	9.062 &	9.062 &	9.073 & 9.139 &	9.139 \\ [1ex] 
 \hline
$q$ & 0.526 & 0.578 & 0.631 & 0.684 & 0.736 & 0.789 & 0.842 & 0.894 & 0.947 & 1 \\ [0.5ex] 
 \hline\hline
 $D$ & 0.208 & 0.204 & 0.199 & 0.194 & 0.189 & 0.184 & 0.178 & 0.171 & 0.164 & 0.156 \\ 
 $S$ & 0.350 & 0.346 & 0.344 & 0.340 & 0.339 & 0.337 & 0.334 & 0.332 & \textit{0.330} & \textbf{0.328} \\
 $r_{.25}$ & \textbf{4.041} &	\textbf{4.041} & \textbf{4.041} &	\textit{3.990} & \textit{3.990} & 3.939 & 3.939 & 3.939 & 3.939 & 3.939 \\
 $r_{.50}$ & 5.703 &	5.686 &	5.703 &	5.703 &	5.703 &	5.703 &	5.703 &	5.703 & 5.667 &	5.650 \\
 $r_{.75}$ & 7.715 &	7.766 &	7.855 &	7.804 &	7.804 &	7.905 &	7.905 &	7.968 & 7.867 &	7.778 \\
 $r_{.90}$ & 9.215 &	9.193 &	9.269 &	9.269 &	9.355 &	9.387 &	9.418 &	9.471 & \textit{9.458} &	\textbf{9.491} \\ [1ex]  
 \hline\hline
\end{tabular}
\caption{ Summary of results for link strength perturbation experiment on a bistable lattice described in Section \ref{section:pertk-bistable}. The best values of $D, S, r_{.25}, r_{.50}, r_{.75}$ and $r_{.90}$ are \textbf{bolded}. The second best are \textit{italicized}.   }
\label{table:pertk-bistable}
\end{table}

%\begin{figure}[htpb]
%\centering
%    \subfloat[]{
%        \includegraphics[width=0.33\linewidth, clip=true, trim={40 0 60 0}]{dissipation_pertk_bi.eps}
%    }
%    \subfloat[]{
%     \includegraphics[width=0.33\linewidth, clip=true, trim={40 0 60 0}]{severity_pertk_bi.eps}
%    } 
%    \subfloat[]{
%        \includegraphics[width=0.33\linewidth, clip=true, trim={40 0 60 0}]{radii_pertk_bi.eps}
%    }
%    \caption{Results of perturbing the stiffnesses of bistable links with  $e_w=.1775$ and mean $\alpha=.3577$, hence $E_{bi} = .4645$ on average, according to discussion in Section \ref{section:pertk-bistable}. This results in the same energy release per destruction of a compound link as the elasto-brittle links studied throughout most of this paper (Sections \ref{section:pertnodes}-\ref{section:combinationofeffects}). (a) Dissipation vs. perturbation magnitude $q$; (b) severity vs. perturbation magnitude $q$; (c) $r_{.25}, r_{.50}, r_{.75}$ and $r_{.90}$ vs. perturbation magnitude $q$.}%%
%\label{fig:pertkbistable}
%\end{figure}

\section{Conclusions}
We have performed numerical experiments on each of four damageable lattices: the triangular lattice, square lattice, hexagonal (``honeycomb") lattice, and aperiodic Penrose lattice. The Penrose and hexagonal lattices exhibit superior resilience to impacts. This is likely caused by the aperiodicity of the underlying lattice structure. 

The broken symmetry of the lattice structure is responsible for distributing kinetic energy more evenly across the structure; the broken symmetry around each node prevents propagation of damage along only one direction. Following this line of reasoning, we have performed a series of numerical experiments to break the symmetry, and thereby improve the resilience, of the periodic triangular lattice. 

\begin{table}[htpb]
\centering
\begin{tabular}{||c c c c c c c c||} 
 \hline
intervention & q & $\frac{\Delta D}{D_0}$ & $\frac{\Delta S}{S_0}$ & $\frac{\Delta r_{.25}}{r_{.25,0}}$ & $\frac{\Delta r_{.50}}{r_{.50,0}}$ & $\frac{\Delta r_{.75}}{r_{.75,0}}$ & $\frac{\Delta r_{.90}}{r_{.90,0}}$ \\ [.5ex] \hline\hline
$x_{i}$  & 0.236 & 0.175 & -0.162 & -0.0024 & 0.0126 & -0.039 & -0.0256 \\
$k_{ij}$  & 1 & -0.103 & \textit{-0.204} & 0.091 & 0.087 & \textit{0.047} & \textbf{0.069} \\
\% removed  & 7.8 & 0.123 & -0.191 & 0.108 & 0.061 & -0.041 & -0.030 \\
$x_{i},k_{ij}$, \% removed  & 0.25, 0.50, 8 & 0.086 & \textbf{-0.228} & -0.016 & 0.052 & 0.003 & \textbf{0.069} \\
bistable  & -- & \textbf{1.422} & 0.021 & \textbf{0.257} &\textbf{ 0.141} & \textbf{0.067} & 0.024 \\
$k_{s,ij},k_{w,ij}$; bistable  & 1 & \textit{0.608} & -0.127 & \textit{0.225} & \textit{0.101} & 0.027 & \textit{0.046} \\ [1ex]
\hline \hline
\end{tabular}
\caption{ Summary of design interventions and relative change as compared with unmodified elasto-brittle triangular lattice ($D_0,S_0,r_{.25,0},r_{.50,0},r_{.75,0}, r_{.90,0}$) for selected values of $q$. In each column, the best value is \textbf{bolded}, and the second best is \textit{italicized}. }
\label{table:summary}
\end{table}

\paragraph{Results.} The results are summarized as follows.
\begin{itemize}
    \item Randomly perturbing node locations leads to increases in energy dissipation, up to about 17\% at $q=0.236$, while severity is reduced by about 16\% at the same value, and there are small, mixed effects on the radii of dissipation. See Table \ref{table:pertnodes}. Generally, perturbing node locations tends to promote increased energy dissipation and reductions in local damage concentration ($S$) with minimal effects on the size of the damaged region. 
    \item Randomly perturbing link strength with $q=1$ leads to a roughly 10\% decrease in dissipated energy, while severity is reduced by about 20\% at the same value. Perturbing the link strength leads to relatively large, positive changes in each of the radii corresponding to energy dissipation, from 5-10\%. See Table \ref{table:pertk}. Generally, perturbing link strength tends to increase the size of the damaged region and reduce the severity $S$, possibly while dissipating less energy. The intended application of the material will inform whether the decreased energy dissipation is compensated by a large reduction in severity.
    \item Randomly removing links and reallocating their mass leads to a roughly 10\% increase in energy dissipation when around 8\% of links are randomly removed. There are mixed effects for radii of dissipation, with slight (about 15\%) increases for $r_{.25}$ and $r_{.50}$. See Table \ref{table:removal}. In contrast with perturbations of the nodes or the link strength, removing links tends to promote reductions in $S$ while promoting greater dissipation of energy. Effects on the radii of dissipation are mixed.
    \item When all three forms of randomization considered here are applied at values that are optimal for each individually, reductions in the severity of structural damage are greater than any of the individual effects alone, while increases in energy dissipation are somewhat moderated compared to the randomization of node positions alone. This simple attempt at ``hybridizing" the approaches presented in this work suggests the possibility of optimization by multiple kinds of induced randomness. See Section \ref{section:randomization:combinationofeffects}.
    \item The use of sacrificial elements are well-studied for their damage delocalizing properties \cite{slepyan-ayzenberg-stepanenko-JMPS-2004,cherkaev-et-al-JMMPS-2005,cherkaev-at-al-MOM-2006,cherkaev-leelavanichkul-IJODM-2012,nadkarni-et-al-PRE-2014}. Here, the bistable lattice, without any randomization, already produces very large increases in energy dissipation, 142\% above the control lattice. The radii of dissipation are also greatly increased. However, $S$ is slightly increased, which is undesirable.
    \item The bistable lattice is combined with the symmetry-breaking ideas of the present work by perturbing the stiffness of the sacrificial and waiting links across the lattice. Such a lattice may benefit from randomization: for $q=1$, the dissipation of energy is 60.8\% greater than for the control lattice, and $S$ is reduced by 12.7\%. The radii of dissipation are also increased nearly as much as in the bistable lattice before randomization. In general, randomization of the bistable lattice leads to trade-offs between energy dissipation and severity of damage. See Section \ref{section:pertk-bistable}, Table \ref{table:pertk-bistable}, and Table \ref{table:summary}.
\end{itemize}

\paragraph{Discussion.} Strategic symmetry-breaking improves several markers for resilience of protective structures, such as energy dissipation, and our simple measures of connectedness and spread. This is achieved either geometrically, by constructing the underlying lattice so that it lacks symmetry and natural directions of propagation, or indirectly, by perturbing the material properties of the links used to construct a lattice that is otherwise symmetric about each of its nodes. The introduction of this asymmetry about each node leads to a lack of preferred direction for propagation of breakage: impacts that would otherwise lead to a crack in a single direction cause many disconnected breaks instead.

These findings are somewhat counterintuitive: while much research is dedicated to improving the precision of engineering techniques, we find that it is possible to achieve superior design not by completely eliminating defects, but by allowing an optimal proportion of defects. However, they are well-aligned with recent work on the use of disorder to improve resilience under load \cite{fulco-et-al-arxiv-2024}, and a broader literature on the use of inhomogeneities to arrest or modify crack development \cite{xu-etal-MSMSE-1997,mogilevskaya-et-al-IJOF-2009,damaskinskaya-et-al-POSS-2018,cavuoto-et-al-IJSS-2022}. This paper, along with that of \cite{fulco-et-al-arxiv-2024}, answers in the negative the question raised by Huang \textit{et al}. \cite{huang-et-al-PMS-2015}: ``is a homogeneous reinforcement distribution optimal?"

The presented results are immediately relevant to the design of structures that need to be able to resist damage due to impacts and remain functional, offering a potentially useful new design principle. 

\paragraph{Future work and open questions.} There are numerous future directions that extend the present work.
\begin{itemize}
    \item We have proposed several markers for resilience. Some of the markers of resilience considered here can be optimized by appropriately choosing the level of perturbation. However, it is not possible in general to optimize several markers simultaneously: a single marker for resilience should eventually be determined. Then a wide variety of optimization techniques become available, including machine learning.
    \item We have not considered the effects of randomization on the effective stress tensor of the lattice. Elsewhere \cite{cherkaev-ryvkin-AAM-2019,cherkaev-ryvkin-AAM-2019b}, the preservation of elastic properties after impact was considered. It is possible that lattice topologies and randomization techniques may introduce trade-offs between desirable elastic properties and desirable impact resilience. Future work should consider the elastic properties as well.
    \item While the theory of crack propagation in continua is well-established, analogous theories for discrete lattices are less developed. More generally, it is not clear how to predict the damage profile at a fixed time (nor the evolving damage profile) from first principles. However, it is expected that the dynamics of the discrete lattice and that of a continuum should converge as the node spacing vanishes. A rigorous theoretical framework is yet to be established.
    \item We have considered only a subset of possible design interventions, and only a range of parameters, for the triangular lattice. Other lattice geometries have been considered briefly. It is possible that some design interventions that greatly benefit one type of lattice are of only marginal benefit to another. A fuller exploration of this dependence is desirable.
    \item Since an impact-resistant lattice should be resistant at all angles of impact, future work should also consider asymmetric impulses intended to simulate off-normal impact.
    \item More sophisticated structures should be considered: in particular, layered media and lattices with non-local connectivity.
\end{itemize}

%Future work on this topic will be focused on further characterizing the criterion for resilience, enabling the formulation of a true structural optimization problem; further numerical experiments on other lattice topologies, possibly with connections between next-nearest-neighbors, and with other impact parameters or design constraints; layered media; and the mathematical description of the crack. Such a mathematical description of the damage profile is highly desirable, while at the same time not being well-suited to the same analysis as cracks within continua. In the absence of a mathematical theory of cracks in discrete lattices, structures that are maximally resilient (according to a given criterion) can only possibly be determined via brute force methods, such as trial and error, or machine learning--- and, then, only as local optima.

\section*{Acknowledgments}
TPE gratefully acknowledges support from the National Science Foundation through DMS-213619.

%\printbibliography
\bibliographystyle{plain}
\bibliography{resilientstructures}

\begin{thebibliography}{10}

\bibitem{ajdari-et-al-IJSS-2012}
Amin Ajdari, Babak~Haghpanah Jahromi, Jim Papadopoulos, Hamid Nayeb-Hashemi, and Ashkan Vaziri.
\newblock Hierarchical honeycombs with tailorable properties.
\newblock {\em International Journal of Solids and Structures}, 49(11):1413--1419, 2012.

\bibitem{braun-fernandez-saez-2016}
M.~Braun and J.~Fernández-Sáez.
\newblock A 2d discrete model with a bilinear softening constitutive law applied to dynamic crack propagation problems.
\newblock {\em International Journal of Fracture}, 197(1), 2016.

\bibitem{caccese-et-al-CS-2013}
Vincent Caccese, James~R. Ferguson, and Michael~A. Edgecomb.
\newblock Optimal design of honeycomb material used to mitigate head impact.
\newblock {\em Composite Structures}, 100:404--412, 2013.

\bibitem{cavuoto-et-al-IJSS-2022}
R.~Cavuoto, P.~Lenarda, D.~Misseroni, M.~Paggi, and D.~Bigoni.
\newblock Failure through crack propagation in components with holes and notches: An experimental assessment of the phase field model.
\newblock {\em International Journal of Solids and Structures}, 257:111798, 2022.
\newblock Special Issue in the honour Dr Stelios Kyriakides.

\bibitem{cherkaev-at-al-MOM-2006}
A.~Cherkaev, V.~Vinogradov, and S.~Leelavanichkul.
\newblock The waves of damage in elastic–plastic lattices with waiting links: Design and simulation.
\newblock {\em Mechanics of Materials}, 38(8):748--756, 2006.
\newblock Advances in Disordered Materials.

\bibitem{cherkaev-et-al-JOPCS-2011}
Andrej Cherkaev, Elena Cherkaev, and Seubpong Leelavanichkul.
\newblock Principles of optimization of structures against an impact.
\newblock {\em Journal of Physics: Conference Series}, 319(1):012021, 2011.

\bibitem{cherkaev-et-al-PRSA-2023-optimal-structures}
Andrej Cherkaev, Elena Cherkaev, and Konstantin Lurie.
\newblock Optimal structures for focusing and energy accumulation: mathematical models and intuition.
\newblock {\em Proceedings of the Royal Society A: Mathematical, Physical and Engineering Sciences}, 479(2277):20220342, 2023.

\bibitem{cherkaev-et-al-JMMPS-2005}
Andrej Cherkaev, Elena Cherkaev, and Leonid Slepyan.
\newblock Transition waves in bistable structures. i. delocalization of damage.
\newblock {\em Journal of the Mechanics and Physics of Solids}, 53(2):383--405, 2005.

\bibitem{cherkaev-leelavanichkul-IJOES-2012}
Andrej Cherkaev and Seubpong Leelavanichkul.
\newblock Approaches to description of damageable lattices dynamics.
\newblock {\em International Journal of Engineering Science}, 58:35--44, 2012.
\newblock Recent advances in Micromechanics.

\bibitem{cherkaev-leelavanichkul-IJODM-2012}
Andrej Cherkaev and Seubpong Leelavanichkul.
\newblock An impact protective structure with bistable links.
\newblock {\em International Journal of Damage Mechanics}, 21(5):697--711, 2012.

\bibitem{cherkaev-ryvkin-AAM-2019}
Andrej Cherkaev and Michael Ryvkin.
\newblock Damage propagation in 2d beam lattices: 1. uncertainty and assumptions.
\newblock {\em Archive of Applied Mechanics}, 89(3):485–501, 2019.

\bibitem{cherkaev-ryvkin-AAM-2019b}
Andrej Cherkaev and Michael Ryvkin.
\newblock Damage propagation in 2d beam lattices: 2. design of an isotropic fault-tolerant lattice.
\newblock {\em Archive of Applied Mechanics}, 89(3):503--519, 2019.

\bibitem{cherkaev-cherkaev-JOE-2003}
Elena Cherkaev and Andrej Cherkaev.
\newblock Principal compliance and robust optimal design.
\newblock {\em Journal of Elasticity}, 72(1):71--98, 2003.

\bibitem{cherkaev-cherkaev-CAS-2008}
Elena Cherkaev and Andrej Cherkaev.
\newblock Minimax optimization problem of structural design.
\newblock {\em Computers \& Structures}, 86(13):1426--1435, 2008.
\newblock Structural Optimization.

\bibitem{damaskinskaya-et-al-POSS-2018}
E.~E. Damaskinskaya, V.~L. Hilarov, I.~A. Panteleev, D.~R. Gafurova, and D.~I. Frolov.
\newblock Statistical regularities of formation of a main crack in a structurally inhomogeneous material under various deformation conditions.
\newblock {\em Physics of the Solid State}, 60(9):1821--1826, 2018.

\bibitem{diaaz-kikuchi-IJNME-1992}
Alejandro~R. Díaaz and Noboru Kikuchi.
\newblock Solutions to shape and topology eigenvalue optimization problems using a homogenization method.
\newblock {\em International Journal for Numerical Methods in Engineering}, 35(7):1487--1502, 1992.

\bibitem{francois-et-al-IJSS-2017}
M.L.M. François, L.~Chen, and M.~Coret.
\newblock Elasticity and symmetry of triangular lattice materials.
\newblock {\em International Journal of Solids and Structures}, 129:18--27, 2017.

\bibitem{fulco-et-al-arxiv-2024}
Sage Fulco, Michal~K. Budzik, Hongyi Xiao, Douglas~J. Durian, and Kevin~T. Turner.
\newblock Disorder enhances the fracture toughness of mechanical metamaterials, 2024.

\bibitem{huang-et-al-PMS-2015}
L.J. Huang, L.~Geng, and H-X. Peng.
\newblock Microstructurally inhomogeneous composites: Is a homogeneous reinforcement distribution optimal?
\newblock {\em Progress in Materials Science}, 71:93--168, 2015.

\bibitem{kochmann-et-al-AMR-2017}
Dennis~M. Kochmann and Katia Bertoldi.
\newblock {Exploiting Microstructural Instabilities in Solids and Structures: From Metamaterials to Structural Transitions}.
\newblock {\em Applied Mechanics Reviews}, 69(5):050801, 10 2017.

\bibitem{lipperman-et-al-JMMS-2009}
Fabian Lipperman, Michael Ryvkin, and Moshe Fuchs.
\newblock Design of crack-resistant two-dimensional periodic cellular materials.
\newblock {\em Journal of Mechanics of Materials and Structures}, 4(3):441--457, 2009.

\bibitem{lipton2025energy}
Robert~P Lipton and Debdeep Bhattacharya.
\newblock Energy balance and damage for dynamic fast crack growth from a nonlocal formulation.
\newblock {\em Journal of Elasticity}, 157(1):1--32, 2025.

\bibitem{melancon-AFM-2022}
David Melancon, Antonio~Elia Forte, Leon~M. Kamp, Benjamin Gorissen, and Katia Bertoldi.
\newblock Inflatable origami: Multimodal deformation via multistability.
\newblock {\em Advanced Functional Materials}, 32(35):2201891, 2022.

\bibitem{mogilevskaya-et-al-IJOF-2009}
Sofia~G. Mogilevskaya, Steven~L. Crouch, Roberto Ballarini, and Henryk~K. Stolarski.
\newblock Interaction between a crack and a circular inhomogeneity with interface stiffness and tension.
\newblock {\em International Journal of Fracture}, 159(2):191--207, 2009.

\bibitem{mohammadi-et-al-JMRT-2023}
Hossein Mohammadi, Zaini Ahmad, Michal Petrů, Saiful~Amri Mazlan, Mohd~Aidy {Faizal Johari}, Hossein Hatami, and Seyed~Saeid {Rahimian Koloor}.
\newblock An insight from nature: honeycomb pattern in advanced structural design for impact energy absorption.
\newblock {\em Journal of Materials Research and Technology}, 22:2862--2887, 2023.

\bibitem{mousanezhad-et-al-IJMS-2014}
D.~Mousanezhad, R.~Ghosh, A.~Ajdari, A.M.S. Hamouda, H.~Nayeb-Hashemi, and A.~Vaziri.
\newblock Impact resistance and energy absorption of regular and functionally graded hexagonal honeycombs with cell wall material strain hardening.
\newblock {\em International Journal of Mechanical Sciences}, 89:413--422, 2014.

\bibitem{nadkarni-et-al-PRE-2014}
Neel Nadkarni, Chiara Daraio, and Dennis~M. Kochmann.
\newblock Dynamics of periodic mechanical structures containing bistable elastic elements: From elastic to solitary wave propagation.
\newblock {\em Phys. Rev. E}, 90:023204, 2014.

\bibitem{ostoja-starzweski-IJSS-1998}
M.~Ostoja-Starzewski.
\newblock Random field models of heterogeneous materials.
\newblock {\em International Journal of Solids and Structures}, 35(19):2429--2455, 1998.

\bibitem{ostoja-starzweski-et-al-EFM-1997}
M.~Ostoja-Starzewski, P.Y. Sheng, and I.~Jasiuk.
\newblock Damage patterns and constitutive response of random matrix-inclusion composites.
\newblock {\em Engineering Fracture Mechanics}, 58(5):581--606, 1997.

\bibitem{ramakrishnan-et-al-JAP-2020}
Vinod Ramakrishnan and M.~J. Frazier.
\newblock {Multistable metamaterial on elastic foundation enables tunable morphology for elastic wave control}.
\newblock {\em Journal of Applied Physics}, 127(22):225104, 06 2020.

\bibitem{slepyan-et-al-JMPS-2005}
Leonid Slepyan, Andrej Cherkaev, and Elena Cherkaev.
\newblock Transition waves in bistable structures. ii. analytical solution: wave speed and energy dissipation.
\newblock {\em Journal of the Mechanics and Physics of Solids}, 53(2):407--436, 2005.

\bibitem{slepyan-ayzenberg-stepanenko-JMPS-2004}
L.I. Slepyan and M.V. Ayzenberg-Stepanenko.
\newblock Localized transition waves in bistable-bond lattices.
\newblock {\em Journal of the Mechanics and Physics of Solids}, 52(7):1447--1479, 2004.

\bibitem{soto-IJOC-2004}
C~A Soto.
\newblock Structural topology optimization for crashworthiness.
\newblock {\em International Journal of Crashworthiness}, 9(3):277--283, 2004.

\bibitem{vainchtein-et-al-PRB-2009}
Anna Vainchtein and Erik~S. Van~Vleck.
\newblock Nucleation and propagation of phase mixtures in a bistable chain.
\newblock {\em Phys. Rev. B}, 79:144123, 2009.

\bibitem{xu-etal-MSMSE-1997}
X-P Xu, A~Needleman, and Farid~F Abraham.
\newblock Effect of inhomogeneities on dynamic crack growth in an elastic solid.
\newblock {\em Modelling and Simulation in Materials Science and Engineering}, 5(5):489, 1997.

\bibitem{yoshida-PLA-1990}
Haruo Yoshida.
\newblock Construction of higher order symplectic integrators.
\newblock {\em Physics letters A}, 150(5-7):262--268, 1990.

\bibitem{zareei-et-al-PNAS-2020}
Ahmad Zareei, Bolei Deng, and Katia Bertoldi.
\newblock Harnessing transition waves to realize deployable structures.
\newblock {\em Proceedings of the National Academy of Sciences}, 117(8):4015--4020, 2020.

\end{thebibliography}

\end{document}